\documentclass[12pt]{article}
\usepackage{amsmath}
\usepackage{amssymb}
\usepackage{amsfonts}
\usepackage{amsthm,systeme}
\usepackage[usenames]{color}
\usepackage[skip=1ex]{caption}
\definecolor{AV}{rgb}{0.65,0.0,0}
\definecolor{GC}{rgb}{0,0.0,0.65}
\definecolor{WS}{rgb}{0,0.65,0}
\usepackage{graphics,epstopdf}
\usepackage[dvips]{graphicx}
\usepackage[margin=1.5cm]{geometry}
\usepackage{graphicx,subcaption,lipsum}
\usepackage[skip=-80pt]{subcaption}
\usepackage{subcaption}
\usepackage{tikz} 
\usepackage{multirow}
\usepackage{tabularx}
\newcolumntype{Y}{>{\centering\arraybackslash}X}
\usepackage{bbding}

\usepackage[
      colorlinks=true,
      linkcolor=blue,
      urlcolor=blue,
      filecolor=blue,
      citecolor=blue,
      pdfstartview=FitV,
      pdftitle={},
      pdfauthor={},
      pdfsubject={},
      pdfkeywords={},
      pdfpagemode=None,
      bookmarksopen=true
]{hyperref}
\usepackage{float}
\usepackage{breqn}
\usepackage{subcaption}
\usepackage[section]{placeins}
\usepackage[nodisplayskipstretch]{setspace}
\newcommand{\beqs}{\begin{eqnarray}}
\newcommand{\eeqs}{\end{eqnarray}}

\usepackage{caption}
\begin{document}

\thispagestyle{empty}

\hfill{}

\hfill{}

\hfill{}

\vspace{32pt}

\begin{center}

\textbf{\Large Epicyclic motion of charged particles around a weakly magnetized Kiselev black hole}

\vspace{48pt}

\textbf{Marina-Aura Dariescu}\footnote{E-mail: \texttt{marina@uaic.ro}} and
\textbf{Vitalie Lungu}\footnote{Corresponding author e-mail: \texttt{vitalie.lungu@student.uaic.ro }}

\vspace*{0.2cm}

\textit{Faculty of Physics, ``Alexandru Ioan Cuza" University of Iasi}\\[0pt]
\textit{11 Bd. Carol I, Iasi, 700506, Romania}\\[.5em]

\end{center}

\vspace{30pt}

\begin{abstract}
We investigate the motion of charged particles evolving around a magnetized Kiselev black hole, in the weak magnetic field approximation. The effective potential allows us to study the bound motion and the stable circular orbits. We analyse the impact of combined quintessence and magnetic fields on the epicyclic frequencies.
Finally, we examine the periapsis shift and gravitational Larmor precession pointing out differences from the Ernst or Kiselev spacetimes.
\end{abstract}

\begin{flushleft}
{\it Keywords}: Kiselev black hole, Ernst solution, Quintessence, Circular orbits, Epicyclic frequencies.
\end{flushleft}

\setcounter{footnote}{0}

\baselineskip 1.5em

\newpage

\section{Introduction}

The study of particle dynamics in the vicinity of black holes is a fundamental tool for revealing the connection between theory and observational data \cite{Chandrasekhar}. In realistic astrophysical situations, black holes are not isolated objects being embedded in magnetized plasma and surrounded by various forms of matter that have a strong influence on motion of test particles. In this respect, the presence of the magnetic fields and dark energy play an important role in shaping the dynamics of the accretion disks, relativistic jets and the mechanism of high energy radiation emitted near black holes \cite{Blandford:1977ds}. 

Observational evidences indicate that supermassive black holes in active galactic nuclei are immersed in magnetic fields of order $10^{4}\,\mathrm{G}$ \cite{Daly:2019srb}, while stellar-mass black holes may be associated with fields as strong as $10^{8}\,\mathrm{G}$ \cite{Piotrovich:2020ooz, Beck:2013bxa}. Even though these magnetic fields are too weak to modify the spacetime geometry, they strongly influence the motion of charged particles via the Lorentz interaction. Thus, the test particle trajectories experience substantial changes, as for example the modification of the innermost stable circular orbit (ISCO) or value and sign of the periapsis shift \cite{Frolov:2010mi}. 

Theoretically, for describing the geometry of a Schwarzschild black hole embedded in an external magnetic field, the solution of Einstein's equations derived by Ernst is the most popular one \cite{Ernst:1976mzr}. This geometry has provided a fertile ground for exploring the dynamics of charged particles in magnetized backgrounds and revealed a rich variety of orbital behavior with no equivalent in Schwarzschild case. For example, due to the Lorentz interaction, the trajectories may have curly or trochoidal shape, experiencing modifications of epicyclic frequencies and exhibiting novel types of orbital precession \cite{Frolov:2010mi, Lim:2015oha}. These effects have been extensively analyzed, using both analytically and numerically methods, for a variety of magnetized black hole spacetimes. See, for examples the references \cite{Kolos:2015iva}-\cite{Taylor:2025ixw}.

On the other hand, a considerable attention has been devoted to the influence of dark energy-like fluids on black hole physics and one the most popular model is the quintessence. This has been modelled as an anisotropic fluid characterized by an equation of state parameter $w$ in the range $-1 \leq w \leq -1/3$ in order to induce the Universe acceleration \cite{SupernovaSearchTeam:1998fmf, SupernovaCosmologyProject:1998vns}. About 20 years ago, Kiselev has obtained an exact, static, spherically symmetric solution of Einstein's equations which describe a black hole surrounded by a quintessential fluid \cite{Kiselev:2002dx}. This solution generalizes the Schwarzschild metric by introducing additional terms in the metric function which are controlled by a parameter related to the dark energy density. The large scale structure of the Kiselev geometry is substantially modified compared to the asymptotically flat spacetime. Thus, depending on the value of $w$, the spacetime may admit, besides the black hole horizon, a cosmological one and the conditions for the existence and stability of orbits are significantly altered. Recently, the Kiselev geometry has been reinterpreted as an exact solution in the context of nonlinear electrodynamics \cite{Dariescu:2022kof}. Following Kiselev's seminal work, the geodesic motion, photon trajectories and circular orbits in Kiselev spacetimes have been widely investigated \cite{Fernando:2012ue}-\cite{Dariescu:2025uck} pointing out that the particle trajectories are affected by the presence of quintessence due to the repulsive contribution that plays an important role at large distances.

Despite the extensive literature on magnetized black holes on one hand and on black holes surrounded by quintessence on the other, the combined effects of these two  physically motivated ingredients have been explored only recently \cite{Abdujabbarov:2015pqp, Rayimbaev:2022mrk}. The coexistence of an external magnetic field and a quintessential fluid gives rise to a spacetime that is neither asymptotically flat nor purely vacuum. In particular, the interplay between the Lorentz force, gravitational attraction and quintessence-driven repulsion can lead to new types of critical points in the effective potential, modified stability conditions and novel orbital structures.

Motivated by the recent interest on these topics, in the present work, we investigate the motion of charged test particles in the spacetime of a magnetized Kiselev black hole, focusing on the astrophysically relevant weak magnetic field regime. The geometry we consider is obtained by applying the magnetization technique of the Kiselev solution presented in \cite{Stelea:2018cgm, Stelea:2018elx} which have led to the axially symmetric spacetime derived in \cite{Lungu:2024iob}. Our solution reduces to the Kiselev metric in the absence of a magnetic field and to the Ernst solution when the quintessence parameter vanishes. By working in the weak magnetic field approximation which is astrophysically relevant, the magnetic induction does not modify the spacetime geometry but retains its dynamical influence on charged particles through the Lorentz force.

Our analysis starts with the construction of the effective potential governing the motion of charged particles and a detailed examination of its critical points. In contrast to the Ernst spacetime, our potential vanishes at both the black hole and cosmological horizons and admits off-equatorial saddle points whose existence is entirely due to the presence of quintessence. These saddle points play a crucial role in determining the conditions under which bound motion is possible and define energy thresholds separating trapped trajectories from escape or capture orbits.

A central part of this work is devoted to the study of circular and quasi-circular orbits. We derive the conditions for the existence of stable circular motion in the equatorial plane and analyse how the ISCO radius depends on the model's parameters. In this respect, we identify parameters ranges for which stable circular orbits exist and provide a systematic classification of allowed angular momenta and energies.
Also, we investigate the quasi-harmonic oscillations around stable circular orbits, by computing the radial and latitudinal epicyclic frequencies and exploring the impact of quintessence and magnetic fields on these characteristic frequencies \cite{Aliev:1989wx}. Particular attention is given to the behavior of the frequencies near the ISCO and near the critical radius beyond which no stable orbits exist. These results are especially relevant in the context of high-frequency quasi-periodic oscillations observed in accreting black hole systems \cite{Villegas:2022jzt}.
Finally, we examine relativistic precessional effects, including periapsis precession and gravitational Larmor (nodal) precession pointing out features that have no analogue in Ernst spacetime \cite{Lim:2015oha}.

The paper is organized as follows. In Section 2, we introduce the magnetized Kiselev spacetime and discuss the structure of its horizons and the equations characterizing the motion of charged particles. Section 3 is devoted to the analysis of the effective potential and its critical points. In Sections 4 and 5, we focus on the bound trajectories and circular orbits. A special attention is given to the innermost stable circular orbits and to the parameters ranges for which such orbits exist. Section 6 addresses the important problem of quasi-harmonic oscillations. The analytical expressions of epicyclic frequencies allow us to discuss the periapsis shift and the gravitational Larmor precession. The conclusions and perspectives are presented in Section 7.

Throughout the paper, we use the conventions $G=c=1$.

\section{Magnetized Kiselev solution}

This work is based on the magnetized Kiselev solution derived in \cite{Lungu:2024iob}
\begin{equation}
ds^2= - f (r) \Lambda^2 dt^2 + \frac{\Lambda^2}{f(r)}  dr^2 + \Lambda^2 r^2 d \theta ^2 + \frac{r^2 \sin^2 \theta}{\Lambda^2} d \phi^2   \; ,
\label{metmag}
\end{equation}
with $\Lambda = 1+B_0^2 r^2 \sin^2 \theta$, where $B_0$ is a constant related to the value of the magnetic induction on the axis of symmetry. In the followings, for simplicity, we shall denote the metric function $f(r)$ by $f$.
 The electromagnetic potential
\begin{equation}
A_{\phi} \, = \frac{B_0 r^2 \sin^2 \theta}{\Lambda}
\label{A3}
\end{equation}
is generating a magnetic field which is axially aligned, breaking the spherical symmetry of the spacetime.  
The components of the magnetic field are given by the expressions
\begin{equation}
B_r=-\frac{4B_0 r\sin\theta \cos\theta}{\Lambda } ,\: B_{\phi}=\frac{4B_0r^2\sin^2\theta}{\Lambda}.
\end{equation}

The contribution of the quintessential fluid is contained in the metric function $f$ which has the general expression obtained by Kiselev \cite{Kiselev:2002dx}
\begin{equation}
f= 1 -\frac{2M}{r} - \frac{k}{r^{3 w +1}} \; ,
\label{fKis}
\end{equation}
where $w$ is the equation of state parameter. In the above relation, $k$ is the quintessence parameter,
which is related to the quintessence fluid energy density as:
\begin{equation}
\rho^0  =  - \frac{3kw}{8\pi r^{3(w+1)}}
\label{rho0}
\end{equation}
while the components of the anisotropic pressures are: $p_r^0=-\rho^0$ and
\begin{equation}
p_t^0 \equiv p_{\theta}^0=p_{\phi}^0=-\frac{3(3w+1)kw}{16 \pi r^{3(w+1)}}.
\label{p0}
\end{equation}
In the presence of the magnetic field, the fluid energy density and pressures are modified and their explicit expressions can be found in \cite{Lungu:2024iob}. However, for a weak magnetic field, which is the case considered in the present work, the expressions of energy density and pressures can be approximated to the ones given in (\ref{rho0}) and (\ref{p0}).  
In what it concerns the values of the quintessence parameter $k$, some constraints have been obtained by comparing the theoretical results with data provided by the EHT Collaboration for M87* and SgrA* \cite{EventHorizonTelescope:2019dse}. For example, in \cite{Atamurotov:2022nim}, the authors are assuming that these objects can be modelled as Kerr supermassive black holes surrounded by quintessence. From the shadows observables, they found that the upper limits on the quintessence parameter are: $k M = 0.0358$ for M87*  BH and $k M = 0.011$ for SgrA* BH. The latter result agrees with the recent estimation for a Frolov BH surrounded by quintessence $k M \leq 0.012$ \cite{Gohain:2024piy}.

In what it concerns the equation of state parameter $w$, this should belong to the interval $w \in [ -1 , -1/3]$ in order to induce an accelerated expansion of the Universe. The particular value $w = -2/3$ has been intensively used in theoretical models since, in this case, one has a linear contribution in the metric function and the whole analysis becomes simpler leading to significant analytical results.  However, a special attention should be also given to the value $w=-1$ which is favoured by cosmological data.

Finally, an important parameter which has a strong influence on the particle's trajectory is the magnetic induction $B_0$. As we are going to see, the model's parameters $B_0$ and $k$ compete against each other and one has to find ranges of these parameters for which the effective potential has a minimum value between two maxima and particles with suitable energies can be trapped in bound orbits.

One may notice that, for $B_0=0$, the relation (\ref{metmag}) with (\ref{fKis}) is the usual Kiselev line element, while for $k=0$, we recover the Ernst solution \cite{Ernst:1976mzr}, also known as the Schwarzschild--Melvin solution.

\subsection{The structure of the horizons}

Depending on the values of $k$ and $w$, there can be up to two horizons which are solutions of the equation $f =0$. 
As a physically important example, let us consider $w=- 2/3$.
This value is intensively used in theoretical studies because it leads to a relative simplicity of treatment due to the linear contribution in the metric function which becomes
\begin{equation}
f = 1 -\frac{2M}{r} - kr 
\label{metric}
\end{equation}
The radial coordinate $r$ is in the range $r \in [r_+ , r_-]$, with
\begin{equation}
r_{\pm} = \frac{1\pm \sqrt{1-8kM}}{2k} 
\label{rpm}
\end{equation}
where $r_-$ is the black hole's horizon, while $r_+$ is the so-called cosmological horizon.
The quintessence parameter must be set below a maximum value $k \leq k_* = 1/(8M)$ for which the two horizons coincide.

For $w=-1/3$, there is just one horizon $r_-$, given by
\begin{equation}
r_-=\frac{2M}{1-k}
\label{rminus}
\end{equation}
which is larger than the Schwarzschild horizon.

In the particular case $w=-1$, the two horizons exist only for very small values of $k$ and these are given by 
\begin{equation}
r_-=\frac{2}{\sqrt{3k}}\sin \left[\frac{1}{3}\arcsin (3M\sqrt{3k})\right], \: r_+=\frac{2}{\sqrt{3k}}\sin \left[\frac{\pi}{3}-\frac{1}{3}\arcsin \left(3M\sqrt{3k})\right)\right]
\label{rminusw}
\end{equation}

\begin{figure}[H]
    \centering
    \begin{subfigure}{0.49\textwidth}
        \centering
        \includegraphics[scale=0.5, trim=2cm 7cm 0cm 1cm, clip]{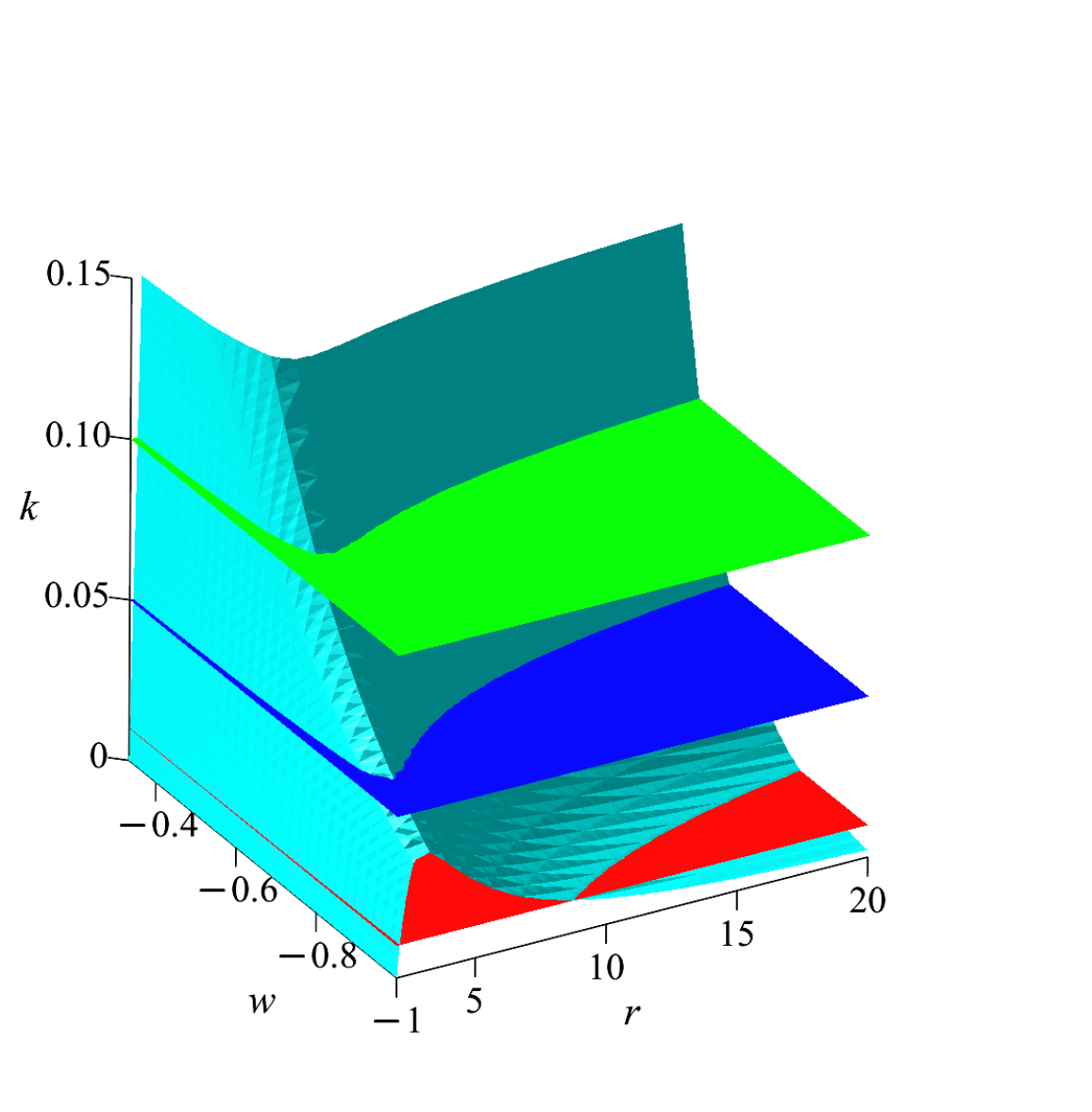}
    \end{subfigure}
    \caption{Plot of the surface (cyan color) $f=0$ as a function of $r$, $w$ and $k$. The horizontal planes corresponds to fixed values of $k$. The fixed values of $k$ are represented by the horizontal planes, the red one corresponds to $k=0.01$, blue to $k=0.05$ and green one to $k=0.1$. Here we used $M=1$. }
    \label{fig:fwfk}
\end{figure}

In the figure \ref{fig:fwfk},  the cyan surface represents the solutions of the equation $f=0$. One may observe that the number of the horizons strongly depends on the values of $k$ and $w$. For a better view, we have represented besides the surface $f=0$, three horizontal planes corresponding to fixed values of $k$. There is always a black hole horizon $r_-$ regardless of the value of $k$ or/and $w$. However, for a fixed $w$, the value of $r_-$ increases with $k$. Besides the black hole horizon, there is a cosmological horizon denoted by $r_+$. As $k$ increases, the cosmological horizon appears for smaller values of $w$. Moreover, for a fixed value of $k$, $r_+$ decreases with $|w|$.

The two horizons coincide if $f=f'=0$. The solutions of this system denoted by $k \equiv k_*$ and $r_*\equiv r_-=r_+$ have the general expressions
\begin{equation}
r_*=\frac{6Mw}{1+3w},  \qquad k_*=-\frac{2M (6Mw)^{3w}}{(1+3w)^{1+3w}}.
\label{rw}
\end{equation}
Several specific examples of $k_*$ and $r_*$ are represented in table \ref{tab:table1}. 

\begin{table}[h!]
\centering
\renewcommand{\arraystretch}{1.5} 
\begin{tabular}{|c|c|c|c|c|}
\hline
$w$ & $-1$ & $-\frac{2}{3}$ & $-\frac{1}{2}$ & $-\frac{1}{3}$\\ 
\hline
$k_*$ & $\frac{1}{27M^2}$ & $\frac{1}{8M}$ & $\frac{1}{9}\sqrt{\frac{6}{M}}$  & $1$ \\
\hline
$r_*$ & $3M$ & $4M$ & $6M$ & $\infty$\\
\hline
\end{tabular}
\caption{The values $k_*$ and $r_*$ for different values of $w$.}
\label{tab:table1}
\end{table}

\subsection{Equations of motion in the weak field approximation}

The motion of a test particle with mass $m_0$ and electric charge $q$ which evolves in this spacetime is described by the following Lagrangian:
\begin{equation}
{\cal L} = \frac{1}{2} \left \lbrace   - f \Lambda^2  \dot{t}^2 + \frac{\Lambda^2}{f}  \dot{r}^2
+ \Lambda^2 r^2 \dot{\theta}^2 + \frac{r^2 \sin^2 \theta}{\Lambda^2} \dot{\phi}^2  \right \rbrace  + \frac{\varepsilon B_0 r^2 \sin^2 \theta}{\Lambda} \dot{\phi},
\label{Lag}
\end{equation}
where $dot$ means the derivatives with respect to the proper time $\tau$ and $\varepsilon = q/m_0$. 

We introduce the dimensionless magnetic parameter $b = \varepsilon B_0$ whose explicit expression is
\[
b = \frac{qB_0GM}{m_0 c^4} = \frac{GM}{c^4} \varepsilon B_0 
\]
which can be directly related to the relative Lorentz force acting on the charged test particle moving around the black hole
with mass $M$ surrounded by quintessence and the external asymptotically
uniform magnetic field $B_0$. 

On the other hand, since the estimated value of the magnetic induction expressed in Gauss is $B_G \sim 10^4$ (G) for supermassive black holes \cite{Daly:2019srb}
and can reach $10^8$ (G) for stellar mass black holes \cite{Piotrovich:2020ooz}, one may assume that these magnetic inductions are too weak compared to the value which can change the
spacetime geometry around the black holes
\[
B = \frac{c^4}{G^{3/2}M} \sim 10^{19} \frac{M_S}{M} \; ({\rm G})
\]
Thus, one may work in the weak field approximation and consider the limit $B_0\to 0$ in $\Lambda$, so that $\Lambda \to 1$. However, the effect of the magnetic field is manifested by means of the Lorentz force acting on charged particles and it has a strong influence on their trajectories. Therefore, we keep the quantity $b = \varepsilon B_0$ finite in the equations of motion.
In this particular case, the Lagrangean (\ref{Lag}) becomes
\begin{equation}
{\cal L} = \frac{1}{2} \left \lbrace   - f  \dot{t}^2 + \frac{\dot{r}^2}{f}
+ r^2 \dot{\theta}^2 + r^2 \sin^2 \theta  \dot{\phi}^2  \right \rbrace  + b r^2 \sin^2 \theta  \dot{\phi},
\label{LagW}
\end{equation}
leading to the constants of motion 
\begin{equation}
E \, = \, f \dot{t} 
\label{Ew}
\end{equation}
and
\begin{equation}
L = r^2 \sin^2 \theta \dot{\phi} + b r^2 \sin^2 \theta \; ,
\label{Lw}
\end{equation}
The corresponding equations of motion:
\begin{eqnarray}
&&
\ddot{r} \,  = \, \frac{2 f -r f^{\prime}}{2} r \dot{\theta}^2 +
\frac{2 f -r f^{\prime}}{2 r^3 \sin^2 \theta}
\left[ L - b r^2 \sin^2 \theta\right]^2 + \frac{2 b f}{r} \left[ L - b r^2 \sin^2 \theta \right] - \frac{f^{\prime}}{2}
\nonumber \\*
& &
\ddot{\theta} \,  = \, 
- \frac{2}{r} \dot{r} \dot{\theta} + \frac{\cot \theta}{r^4 \sin^2 \theta } 
 \left[ L - b r^2 \sin^2 \theta \right]^2 + \frac{2 b \cot \theta}{r^2}   \left[ L - b r^2 \sin^2 \theta \right]
\nonumber \\*
& & \dot{\phi} = \frac{1}{r^2 \sin^2 \theta} \left[ L -b r^2 \sin^2 \theta\right]
\label{weak} 
\end{eqnarray}
can be used to draw the particles trajectories by numerical integration. A detailed analysis and classification of orbits for the general line element (\ref{metmag}) has been developed in \cite{Lungu:2024iob}.

\section{The effective potential and critical points. A three dimensional analysis}

The four-velocity normalization relation
\[
- f \dot{t}^2  + \frac{\dot{r}^2}{f} +  r^2 \dot{\theta}^2  + r^2 \sin^2 \theta \dot{\phi}^2   = -1 \; ,
\]
with the constants of motion (\ref{Ew}) and (\ref{Lw}), leads to the equation
\begin{equation}
 \dot{r}^2 + f r^2 \dot{\theta}^2    = E^2 -V
\label{rdot}
\end{equation}
We have introduced the effective potential:
\begin{equation}
V =  f \left[ 1 + \frac{\left( L - b r^2 \sin^2 \theta \right)^2}{r^2 \sin^2 \theta} \right]
\label{V}
\end{equation}
which is depending on the quintessence parameters $w$ and $k$ and on the strength of the magnetic field $B_0$ by $b = \varepsilon B_0$.
Since the metric function is positive between the horizons, $f \geq 0$, the effective
potential (\ref{V}) is always a positive quantity and is vanishing on the horizons, contrary to Ernst spacetime \cite{Ernst:1976mzr}.
In the followings, we shall consider that the angular momentum $L$ is positive, while $b$ may have either positive or negative values. The other situations corresponding to $L<0$ are equivalent to these ones because the effective potential shows a clear symmetry for
$(L , b) \to (-L , -b)$.
If the magnetic parameter and the angular
momentum have the same signs, the
Lorentz force is repulsive and the particle is pushed outward the black
hole. In the other case where $b$ and $L$ have opposite signs, the
Lorentz force is attracting the charged particle towards the black hole \cite{Frolov:2010mi, Lim:2015oha}. 

In what it concerns the critical points of the potential (\ref{V}), these can be found by imposing the conditions:
\begin{equation}
\partial_r V=  \partial_{\theta}V=0
\end{equation}
which lead to the following equations in terms of the angular momentum $L$
\begin{equation}
(f'r-2f) L^2-2 b  r^3\sin^2\theta f' L+r^3\sin^2\theta\left[( b^2  r^2\sin^2\theta+1) f' +2 b^2 fr \sin^2\theta\right]=0
\label{Lr}
\end{equation}
and
\begin{eqnarray}
f \cos\theta \left[ L^2 -b^2 r^4 \sin^4 \theta \right]=0
\label{Ltheta}
\end{eqnarray}
The real positive roots of the equation (\ref{Lr}) situated between the horizons determine the maxima, minima
and saddle points of the effective potential. 
As expected, in the equatorial plane where the critical points corresponding to the maximum or minimum are located, the left hand side of the equation (\ref{Ltheta}) vanishes. In  order to find the solutions situated outside the equatorial plane, we impose that the equations (\ref{Lr}) and (\ref{Ltheta}) have at least one common solution. Using Cramer's rule one finds that the off-equatorial critical point is a solution of $f^{\prime} =0$, being given by the relation
\begin{equation}
r_s = \left[ - \frac{(3w+1)k}{2M} \right]^{\frac{1}{3w}}
\label{rs}
\end{equation}
One may notice that $r_s$ is independent of the parameters $L$ and $b$.
The relation (\ref{Ltheta}) gives the polar coordinate corresponding of the off-equatorial plane critical points:
\begin{equation}
\sin^2 \theta_s = \pm \frac{L}{b  r_s^2}
\label{thetas}
\end{equation}
where the minus sign is for the case where $L$ and $b$ have opposite signs.
Due to the axial symmetry, there are two points with $\theta_s$, symmetric with respect to the $\pi/2$ plane.
The above relation imposes 
\begin{equation}
|L| \leq	 | b | r_s^2
\label{L*}
\end{equation} 
These off-equatorial critical points correspond to saddle points, as they cannot represent local extrema of the effective potential. This behaviour has no equivalent in Ernst spacetime, where all critical points are confined to the equatorial plane \cite{Ernst:1976mzr, Kolos:2015iva}.

In order to verify that the pair $(r_s , \theta_s )$ which is a critical point of the differentiable real function $V ( r , \theta )$ corresponds to a saddle point, one has to compute the Hessian matrix associated to the potential $V$ . The corresponding determinant 
reads
\begin{equation}
D ( r_s , \theta_s )= \left. \frac{\partial^2 V}{\partial r^2} \frac{\partial^2 V}{\partial \theta^2} - \left(\frac{\partial^2 V}{\partial r \partial \theta}\right)^2 \right|_{r=r_s , \theta = \theta_s } =  \frac{48Mwb^2}{r_s} f_s \cos^2 \theta_s
\label{D}
\end{equation}
with $f_s = f(r=r_s)$.
Since the parameter $w$ is negative and $f_s>0$, one may conclude by saying that the function $D ( r_s , \theta_s )$ is a negative quantity and therefore $( r_s , \theta_s )$ is indeed a saddle point of $V$ \cite{J. Stewart}. The existence of such critical points outside the equatorial plane arises solely due to the presence of quintessence.

\section{Bound motion of charged particles}

\subsection{Conditions for bound orbits to exist}

The existence of saddle points plays a crucial role in defining the trapped motion around the black hole. For a bound orbit to exist, the energy associated to the saddle point must be greater than the energy $E_{\min}^2$ corresponding to the local minimum of the effective potential (\ref{V}) in the equatorial plane, i.e.
\begin{equation}
V_0 =  f \left[ 1 + \frac{\left( L - b r^2 \right)^2}{r^2 } \right]
\label{V0}
\end{equation}
The condition $V_0^{\prime}=0$ gives the radial coordinate of the circular orbit $r=r_c$. If, in addition, one has $V^{\prime \prime} (r=r_c)>0$, the circular orbit located in the equatorial plane is stable and the corresponding energy is: $E_{min}^2 = V_0 ( r=r_c)$.
On the other hand, the energies associated to the saddle points symmetric with respect to the equatorial plane are computed as:
$E_{s1,s2}^2 = V ( r=r_s , \theta = \theta_{s1,s2} )$, where the effective potential is given in (\ref{V}), while $r_s$ and $\theta_{s1,s2}$ are defined in (\ref{rs}) and (\ref{thetas}). 
Thus, the condition for the bound orbit to exist is:
\begin{equation}
E_{min}^2 = V_0<E_{s1,s2}^2
\label{BoundCond}
\end{equation}
where $E_{min}^2$ and $E_{s1,s2}^2$ are strongly depending on the model's parameters $k$, $b$ and $w$, as it can be seen in the figure \ref{fig:energies}.

In the followings, let us consider $L>0$ and give the expressions of $E_{s1}^2$ and $E_{s2}^2$ corresponding to positive and negative values of $b$.
For $b>0$, the energy of the saddle point has the simple expression
\begin{equation}
E_{s1}^2=1-\frac{2M}{r_s}-\frac{k}{r_s^{1+3w}} \equiv f_s,
\label{Es}
\end{equation}
while, for $b<0$, one has
\begin{equation}
E_{s2}^2= f_s (1-4bL)
\label{Es1}
\end{equation}

On the other hand, the constrain (\ref{BoundCond}) can be worked out in order to find ranges of the angular momentum which allow bound orbits. 
As a first case, let us consider $b>0$ for which the energy of the saddle point is $E_{s1}^2$. The angular momentum range corresponding to the condition $V_0<E_{s1}^2$ is: $L_{1}^{-}<L<L_{1}^{+}$, where
\begin{eqnarray}
L_{1}^{\pm}=br^2 \pm r\sqrt{\frac{f_s}{f}-1}
\label{L1}
\end{eqnarray}

In the second case with $b<0$, the angular momentum corresponding to the condition $V_0<E_{s2}^2$ lies in the range: $L_{2}^{-}<L<L_{2}^{+}$, where 
\begin{eqnarray}
L_{2}^{\pm}=\frac{r}{f}\left[br\left(f-2f_s\right)\pm\sqrt{\left(4b^2f_s r^2+f\right)\left(f_s-f\right)}\right]
\label{L2}
\end{eqnarray}
Thus, for a particle to evolve on a bound orbit, its angular momentum should be in the range:
\begin{equation}
L_{1,2}^{-}< L <L_{1,2}^{+}
\label{Lrange}
\end{equation}
In practice, once we fix the model's parameters $b$, $k$ and $w$ and the value of the angular momentum $L$, the relation (\ref{Lrange}) defines the regions of the radial coordinate that host bound orbits.

Finally, let us notice that the relations (\ref{Lrange}) and (\ref{L*}) lead to the maximum value of the radial coordinate of the turning point located in the equatorial plane:
\begin{equation}
r\leq r_*= \left[ - \frac{(3w+1)k}{2M} \right]^{\frac{1}{3w}}
\label{rmax}
\end{equation}
If this condition is satisfied, the expressions of $L_{1,2}^{\pm}$ are real.

Let us underline that $r_*$ can be named as the {\it static radius} (as in \cite{Stuchlik:1999qk}) where the gravitational attraction is compensated by the
quintessence repulsion. This value of the radial coordinate will play an important role in separating the region corresponding to $r<r_*$ where one may have stable orbits, from the one with $r> r_*$ where the particle's trajectories  are always unstable. In our analysis, we shall consider $kM \ll 1$ so that the cosmological horizon is far from the black hole's horizon and from the region where the particle's bound trajectories are confined. This assumption agrees with other theoretical estimates on the upper limits of the quintessence parameter \cite{Gohain:2024piy}.
As an example, let us consider the special case corresponding to $w=-2/3$ for which the two horizons have the analytical expressions given in (\ref{rpm}). For $kM \ll 1$, the expressions of the horizons can be approximated to
\[
r_- \approx 2M \; , \quad r_+ \approx \frac{1}{k}
\]
and one has the relation $r_- < r_* = \sqrt{2M/k} \ll r_+$. 

\subsection{Illustrative examples}

In the followings, let us discuss the dependence of $E_{min}^2$ and $E_{s1,s2}^2$, defined in the previous subsection, on the parameters $k$ and $b$.

\begin{figure}[H]
    \centering
    \begin{subfigure}{0.49\textwidth}
        \centering
        \includegraphics[scale=0.5, trim=2cm 12cm 5cm 1cm, clip]{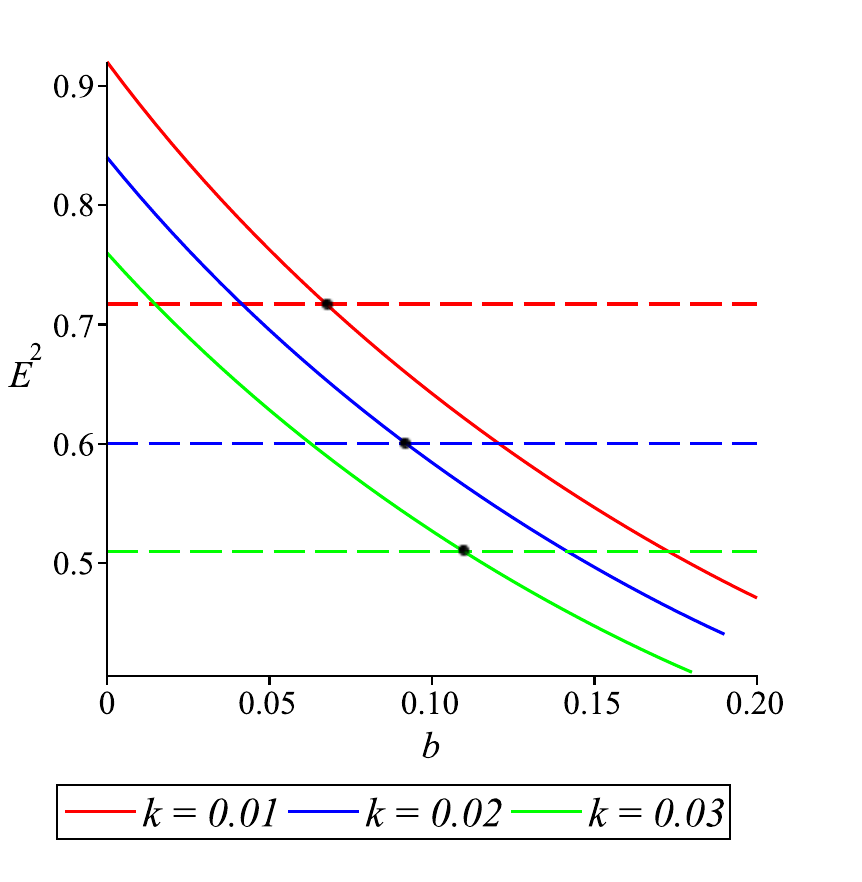}
    \end{subfigure}
    \hfill
    \begin{subfigure}{0.49\textwidth}
        \centering
        \includegraphics[scale=0.5, trim=0cm 12cm 5cm 1cm, clip]{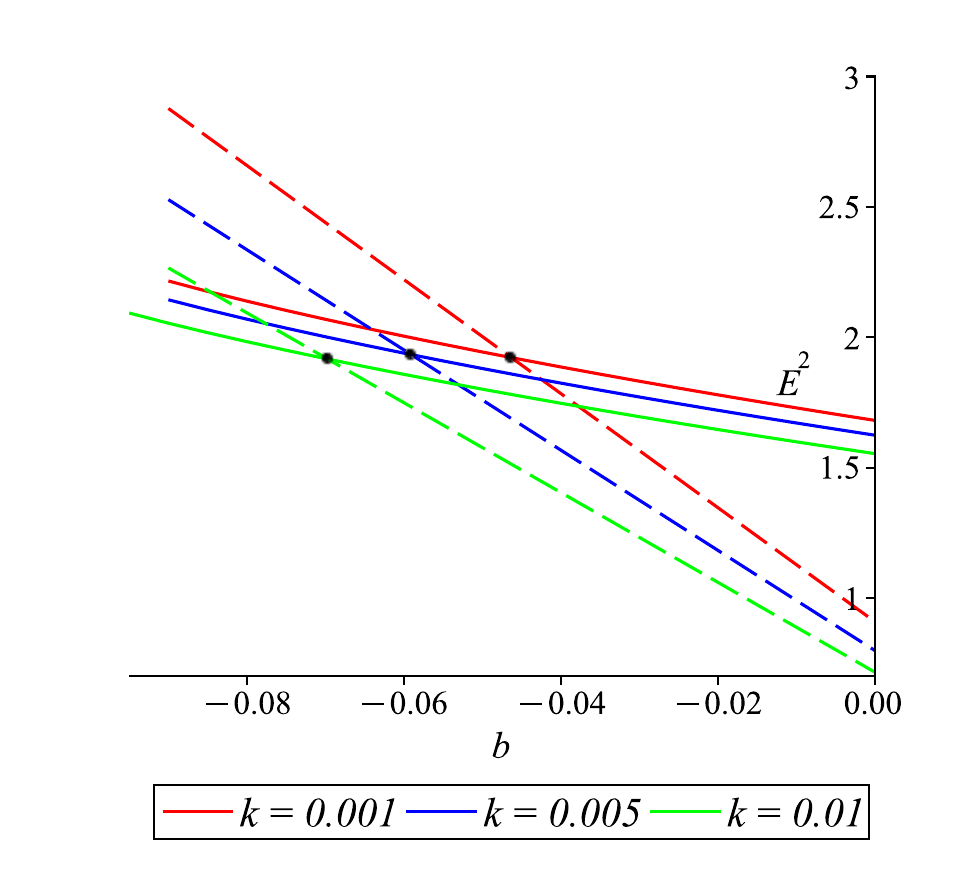}
    \end{subfigure}
    \caption{{\it{Left panel}}. Plot of $E_{\min}^2$ (solid curves) and $E_{s 1}^2$ (dashed horizontal lines) as functions of $b>0$ for different values of $k$. {\it{Right panel}}.  Plot of $E_{\min}^2$ (solid curves) and $E_{s2}^2$ (dashed curves) as functions of $b<0$, for different values of $k$. The black dots represent the critical values $b_{cr}$ for which $E_{\min}^2=E_{s 1, s2}^2$. The other values of the parameters are: $M=1$, $w=-2/3$ and $L=4$.}
    \label{fig:energies}
\end{figure}

In the left panel of figure \ref{fig:energies}, the energies $E_{s1}^2$ are represented by dashed horizontal lines since these are depending only on $k$. On the other hand, the energies $E_{min}^2$ are depending on both $k$ and $b$ and are represented by solid lines. In regions where $E_{\min}^2 > E_{s1}^2$ there are no bound orbits and this happens for very small values of $b$. For a given value of $k$, as $b$ increases, the energy $E_{\min}^2$ decreases. There is a critical value of the magnetic parameter, $b_{cr}$, (corresponding to the black dots) above which one has $E_{\min}^2 < E_{s1}^2$ and bound orbits are allowed. Once the parameter $k$ is increasing, the critical value $b_{cr}$ is also increasing. Thus, as expected, bound orbits are allowed for large values of $b$ and small values of $k$. 

In the second case, corresponding to $b<0$, for which the energy associated to the saddle point is given in (\ref{Es1}) an example is represented in the right panel of the figure \ref{fig:energies}. One may notice that both $E_{s2}^2$ and $E_{min}^2$ are depending on $b$ and $k$ and these are decreasing as $k$ increases. Similarly to the previous case, $|b_{cr}|$ increases with $k$. Compared to the previous case, for $b<0$, the bound orbits are allowed for very small values of $k$ and are  characterized by larger values of the energies. 

\begin{figure}[H]
    \centering
    \begin{subfigure}{0.49\textwidth}
        \centering
        \includegraphics[scale=0.4, trim=2cm 6cm 0cm 1cm, clip]{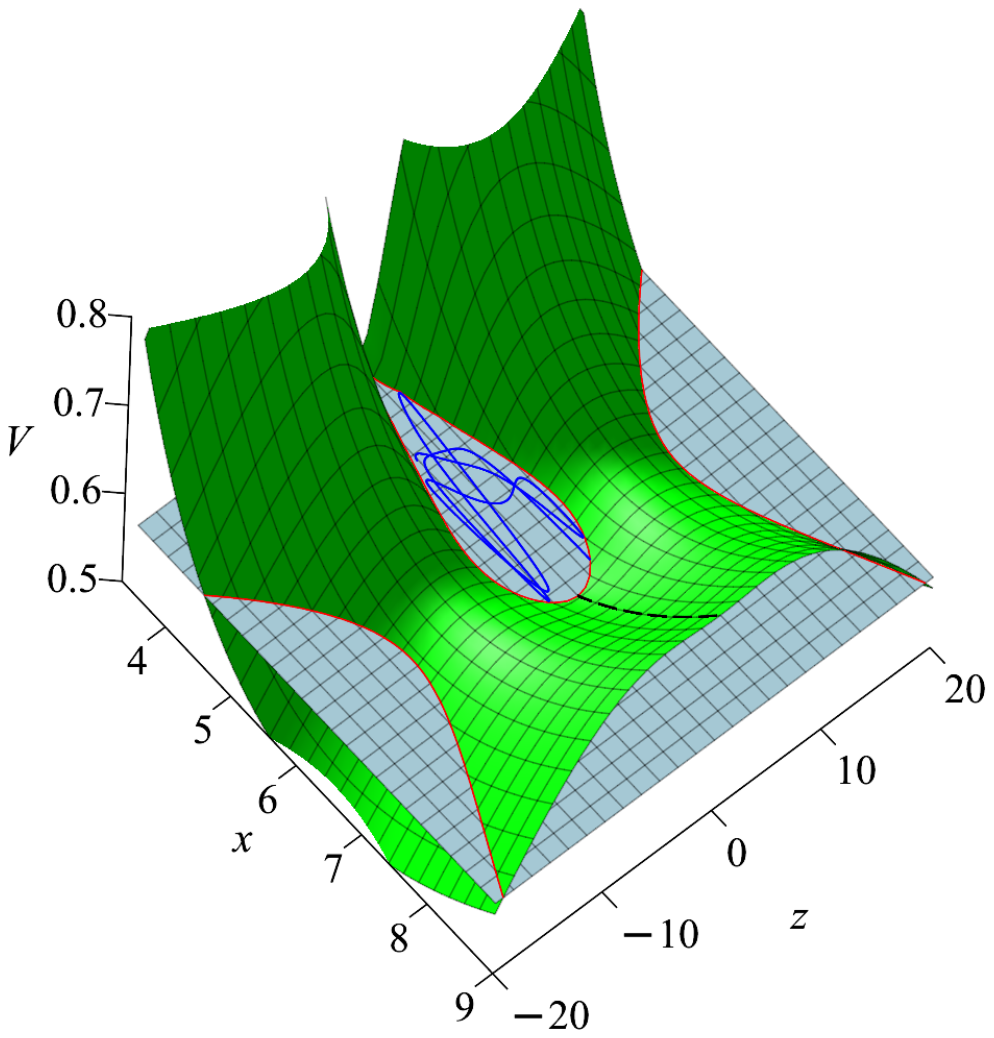}
    \end{subfigure}
    \hfill
    \begin{subfigure}{0.49\textwidth}
        \centering
        \includegraphics[scale=0.6, trim=0cm 12cm 0cm 1cm, clip]{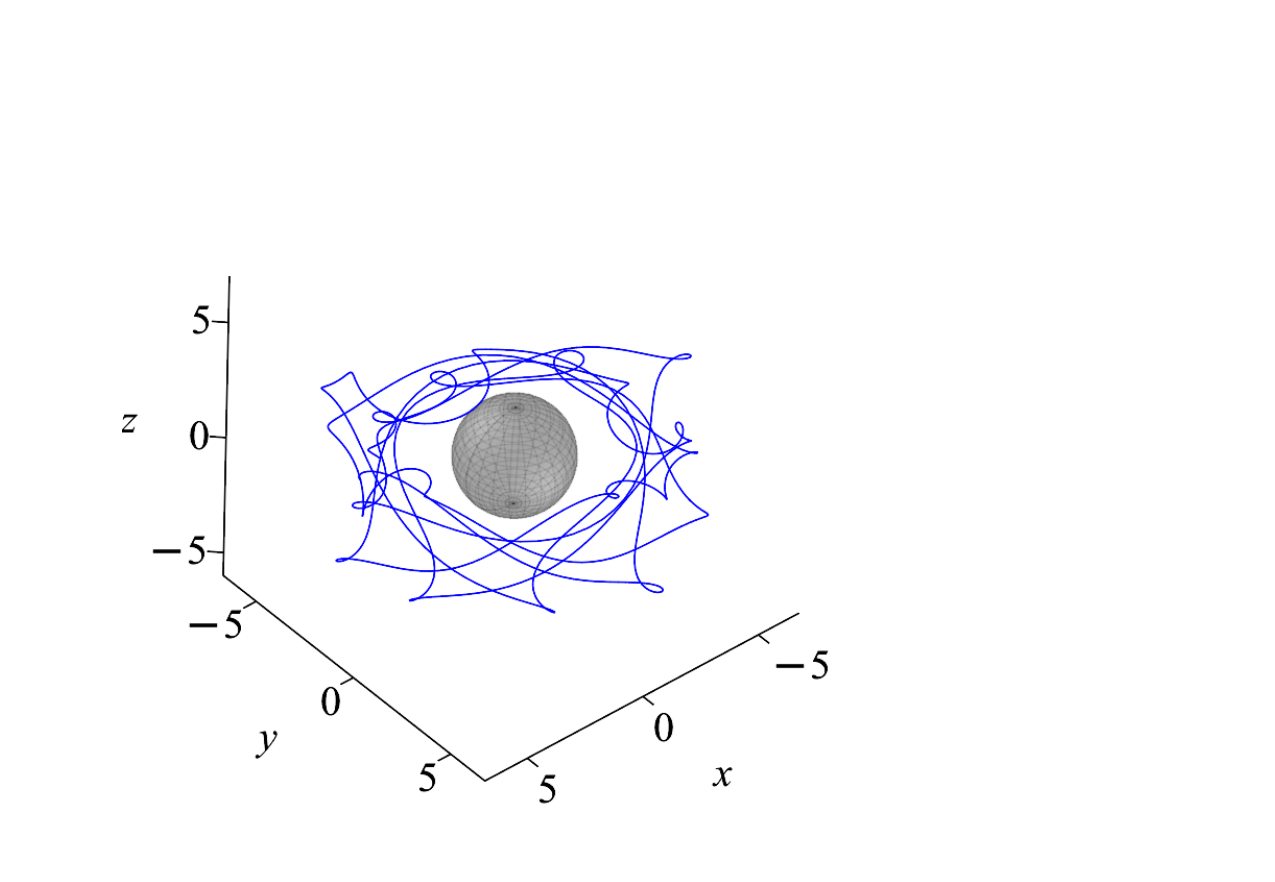}
    \end{subfigure}
    \caption{{\it{Left panel}}. Plot of the effective potential (\ref{V})  and the bound particle trajectory in the $(x,z)-$ plane. The energy $E^2=0.58$ is represented by the light blue horizontal plane. The equipotential curve given by the solutions of the equation $E^2=V$ is represented by the red curves. The equatorial plane corresponding to $z=0$ is represented by the dashed black line. {\it{Right panel}}. 3D plot of the bound trajectory of a particle moving in the potential represented in the left panel. The gray sphere represents the horizon $r_-=2.09$. The values of the parameters are: $M=1$, $w=-2/3$, $k=0.02$, $b=0.10$ and $L=4$. }
    \label{fig:bound}
\end{figure}

\begin{figure}[H]
    \centering
    \begin{subfigure}{0.49\textwidth}
        \centering
        \includegraphics[scale=0.4, trim=2cm 0cm 0cm 1cm, clip]{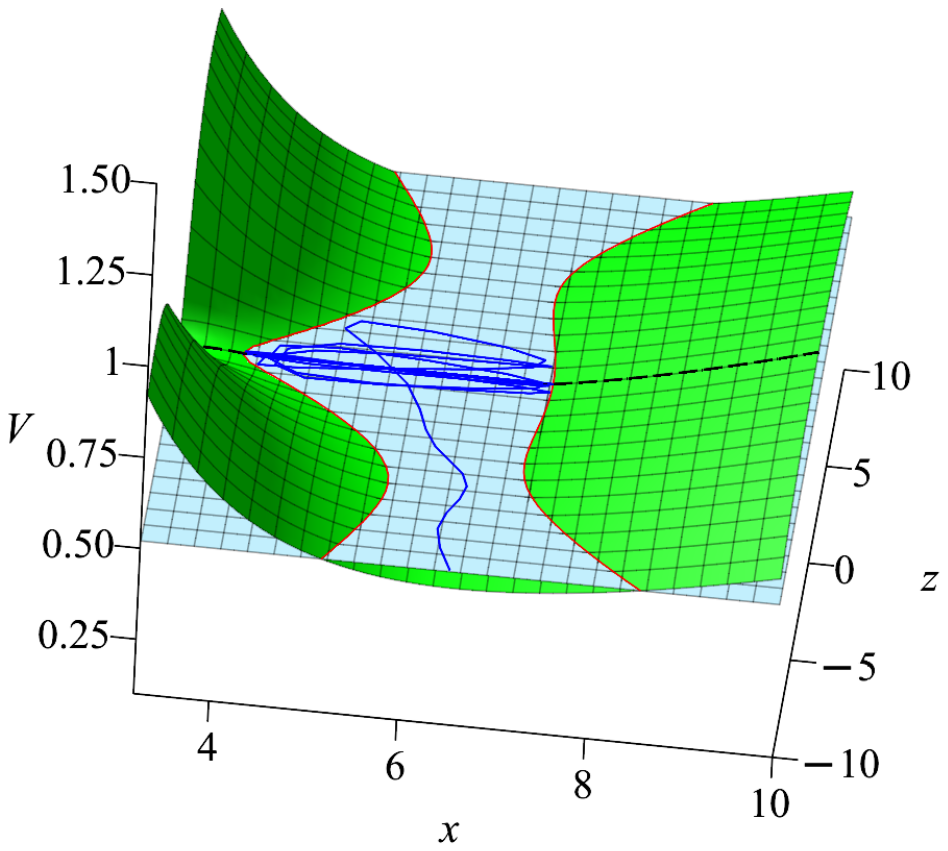}
    \end{subfigure}
        \hfill
    \begin{subfigure}{0.49\textwidth}
        \centering
        \includegraphics[scale=0.5, trim=0cm 4cm 0cm 4cm, clip]{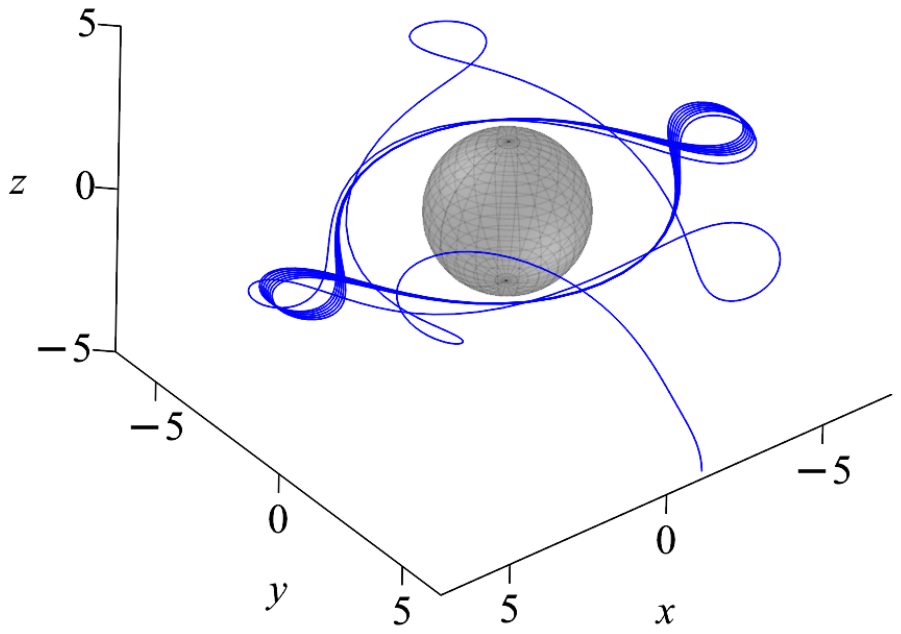}
    \end{subfigure}
    \caption{{\it{Left panel}}. Plot of the effective potential (\ref{V})  and the particle trajectory in the $(x,z)-$ plane. The equipotential curve given by the solutions of the equation $E^2=V$ is represented by the red curve. The equatorial plane corresponding to $z=0$ is represented by the dashed black line. The energy $E^2=0.52$ is represented by the light blue plane. {\it{Right panel}}. 3D plot of the escape trajectory of a particle moving in the potential represented in the left panel. The gray sphere represents the horizon $r_-=2.14$.  The values of the parameters are: $M=1$, $w=-2/3$, $k=0.03$, $b=0.10$ and $L=4$.  }
    \label{fig:escape}
\end{figure}

Finally, let us fix the values of the parameters and give, in the figure \ref{fig:bound}, an example of the effective potential (\ref{V}) and the corresponding bound orbit. One can observe in the left panel of figure \ref{fig:bound} that the particle motion is bounded by the equipotential curve determined by the solutions of the equation $E^2 = V$. In contrast, the figure \ref{fig:escape} illustrates the escape trajectory of a particle whose motion starts in the equatorial plane. However, due to the absence of a Carter constant, the trajectory is not confined to this plane indefinitely, and the particle is oscillating between $\theta_{\min}$ and $\theta_{\max}$. But, since $E_{\min}^2 > E_{s1}^2$, the region of motion is not bounded by an equipotential curve and the particle escapes toward the cosmological horizon.

\begin{figure}[H]
\centering
\begin{subfigure}{0.49\textwidth}
\centering
\includegraphics[scale=0.4, trim=2cm 6cm 0cm 1cm, clip]{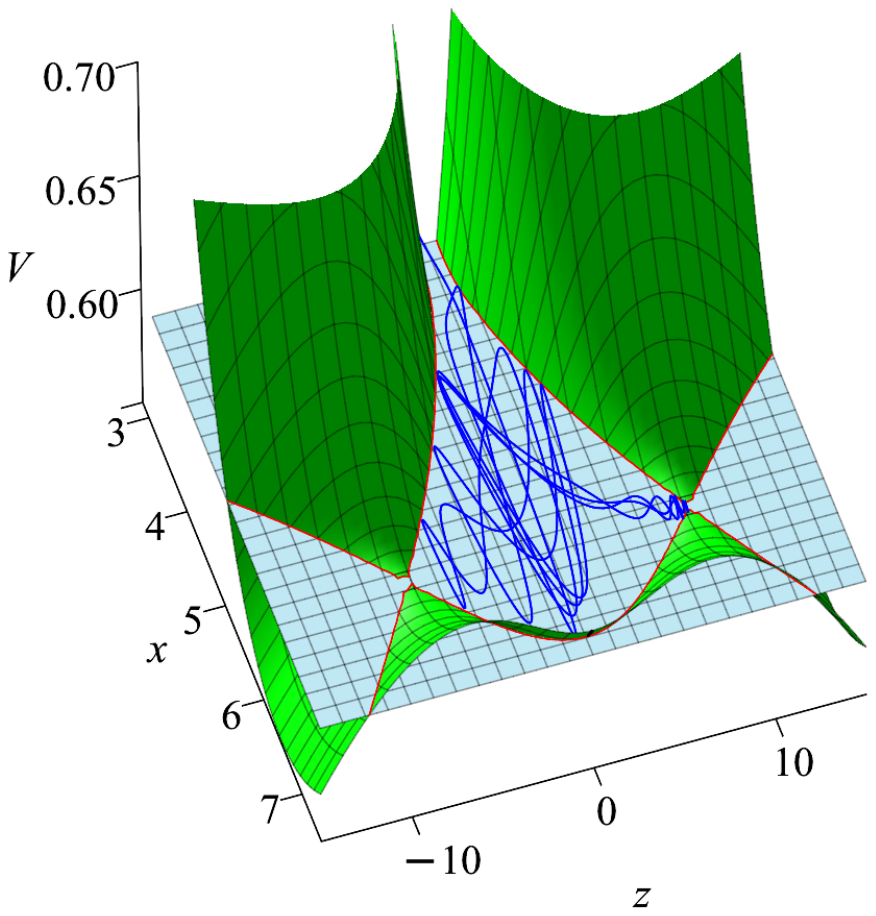}
\end{subfigure}
\hfill
\begin{subfigure}{0.49\textwidth}
\centering
\includegraphics[scale=0.6, trim=2cm 8cm 0cm 1cm, clip]{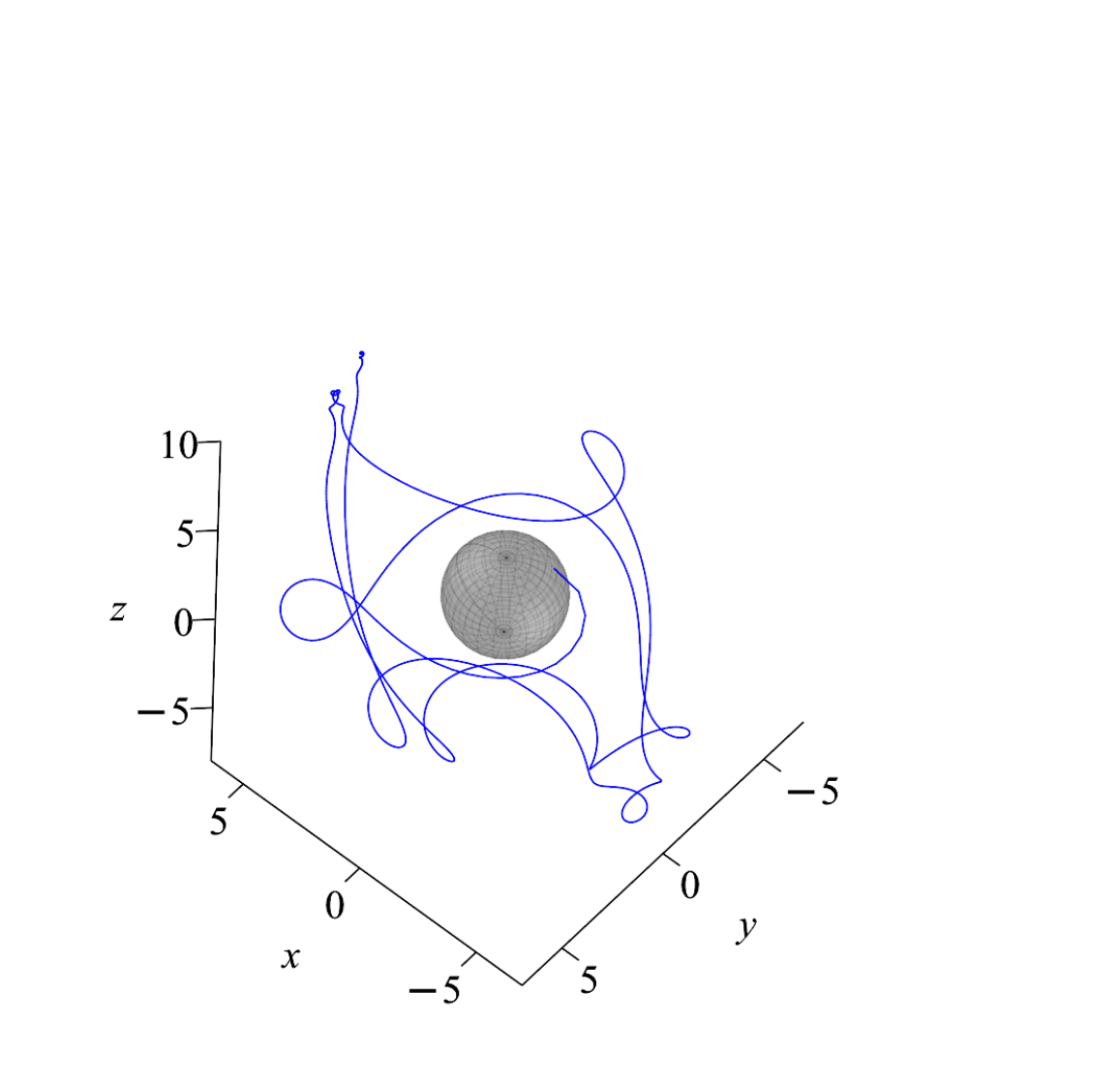}
\end{subfigure}
\caption{{\it{Left panel}}. Plot of the effective potential (\ref{V}) and the capture particle trajectory in the $(x,z)-$ plane. The particle with the energy $E^2=0.60$ starts its motion from the saddle point $(r_s , \theta_s ) = (  10 , 0.68 )$. {\it{Right panel}}. 3D plot of the capture trajectory of a particle moving in the potential represented in the left panel. The gray sphere represents the horizon $r_-=2.09$. The values of the parameters are: $M=1$, $w=-2/3$, $k=0.02$, $b=0.10$ and $L=4$. }
    \label{fig:captures}
\end{figure}

\begin{figure}[H]
\centering
\begin{subfigure}{0.49\textwidth}
\centering
\includegraphics[scale=0.4, trim=2cm 6cm 0cm 1cm, clip]{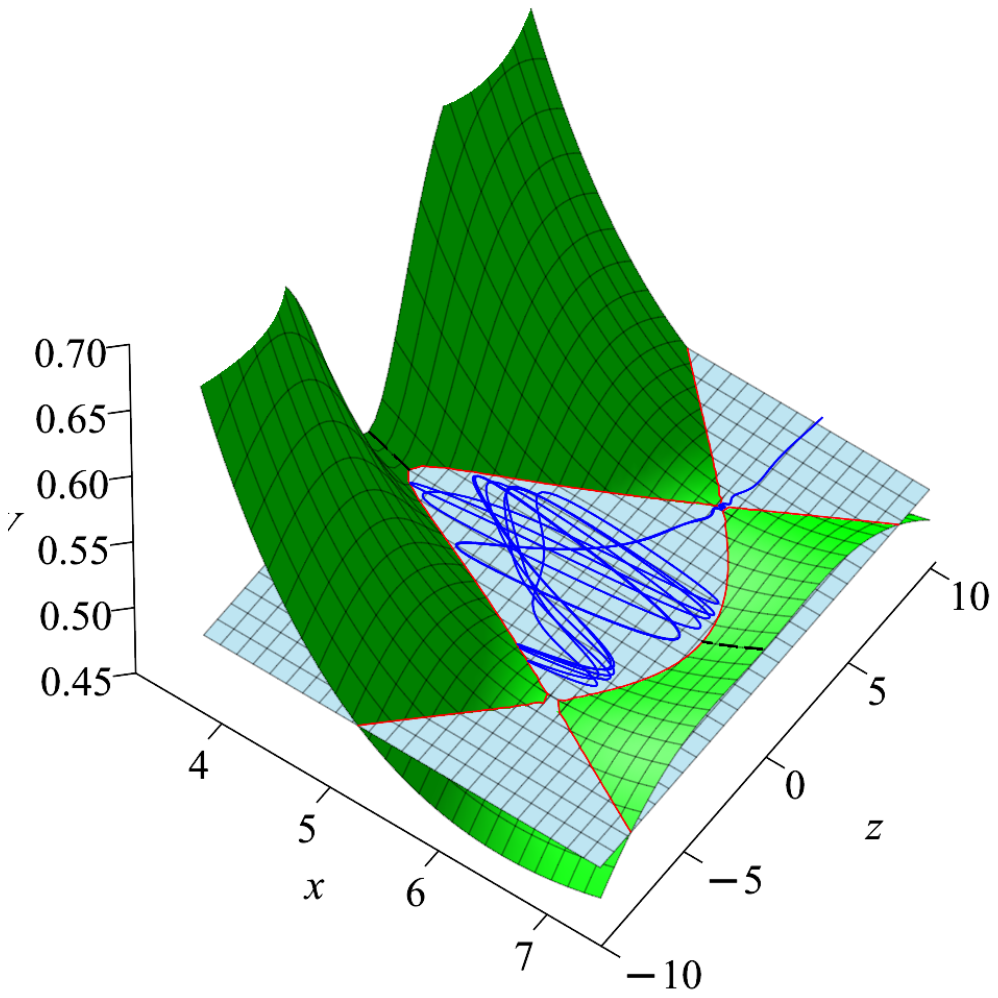}
\end{subfigure}
\hfill
\begin{subfigure}{0.49\textwidth}
\centering
\includegraphics[scale=0.6, trim=2cm 8cm 0cm 1cm, clip]{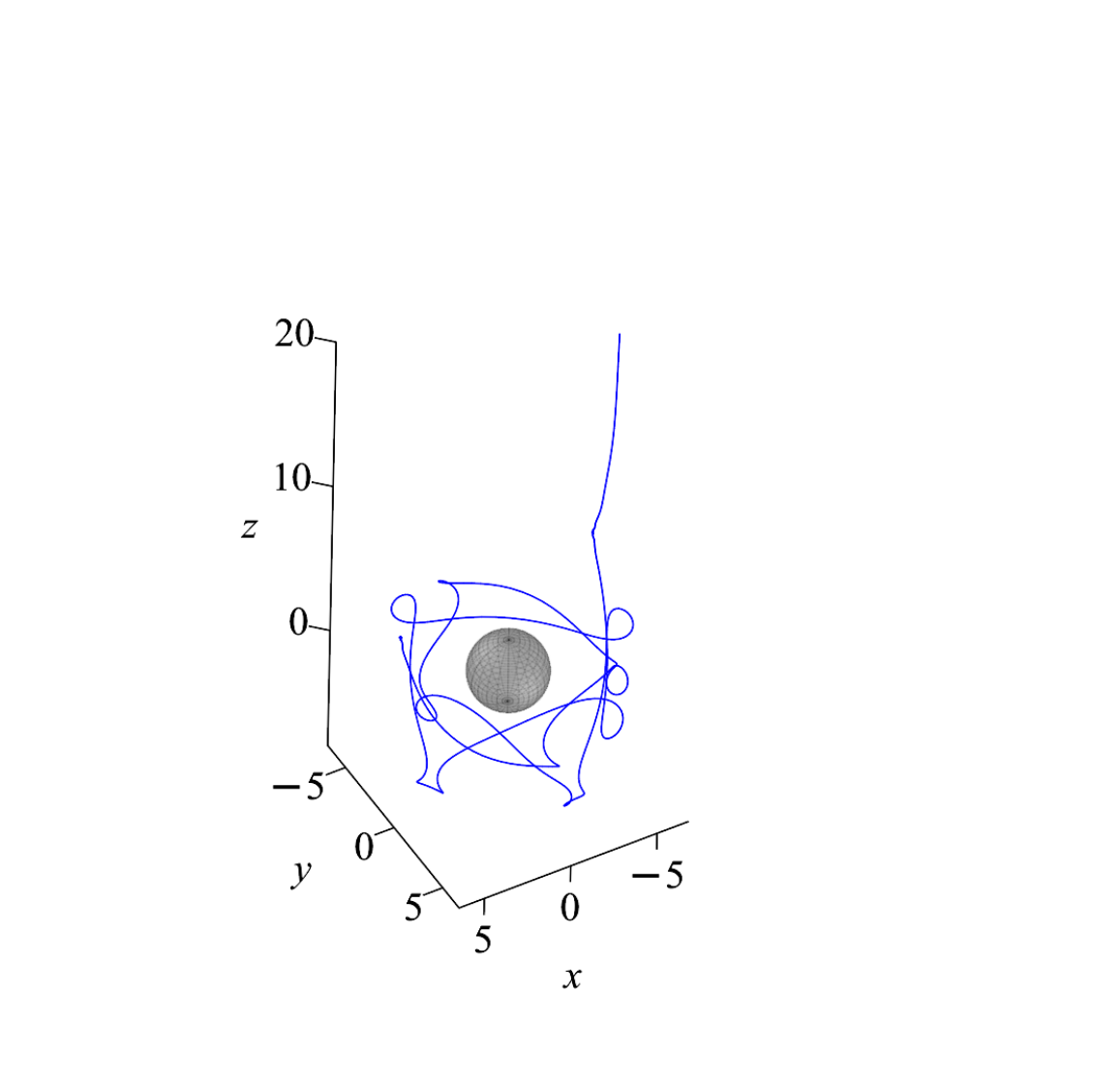}
\end{subfigure}
\caption{{\it{Left panel}}. Plot of the effective potential (\ref{V})  and the escape particle trajectory in the $(x,z)-$ plane. The particle starts its motion from the saddle point $( r_s , \theta_s ) = ( 8.17, 0.89 )$.  The energy $E^2=0.51$ is represented by the light blue horizontal plane. {\it{Right panel}}. 3D plot of the escape trajectory of the particle moving in the potential represented in the left panel. The gray sphere represents the horizon $r_-=2.14$. The values of the parameters are: $M=1$, $w=-2/3$, $k=0.03$, $b=0.10$ and $L=4$. }
    \label{fig:escapes}
\end{figure}

If the particle starts its motion from a saddle point, its trajectory is either a capture one, as in the figure \ref{fig:captures}, or an escape one, as in the figure \ref{fig:escapes}.

Finally, one may notice in the figures that the particle's trajectory can be curly and this happens in cases where $L$ and $b$ have the same sign and $\dot{\phi}$ given by (\ref{weak}) changes sign in $r_0$ given by
\begin{equation}
r_0=\sqrt{\frac{L}{b}}\frac{1}{\sin \theta_0},
\end{equation}
where $r_0$ is between the turning points and $\theta_0 \in [\theta_{\min}, \theta_{\max}]$. For motion confined in the equatorial plane, the radial coordinate $r_0$, where $\dot{\phi}$ changes sign is $r_0=\sqrt{L/b}$.
If $L$ and $b$ have opposite signs, the coordinate $\phi$ is either growing (for $L>0$ and $b<0$) or decreasing (for $L<0$ and $b>0$) during the charged particle motion.
The curly behavior of charged particles trajectories is satisfied for $\sin \theta_0 <1$, i.e.
\begin{equation}
E^2>1-2M\sqrt{b/L}-k (b/L)^{(1+3w)/2}\equiv E_0^2,
\label{E0}
\end{equation}
where one has to impose the condition $E_{min}^2 < E_0^2<E_{s1}^2$.

\section{Circular orbits}

\begin{figure}[H]
    \centering
    \begin{subfigure}{0.49\textwidth}
        \centering
        \includegraphics[scale=0.5, trim=0cm -2cm 0cm -3cm, clip]{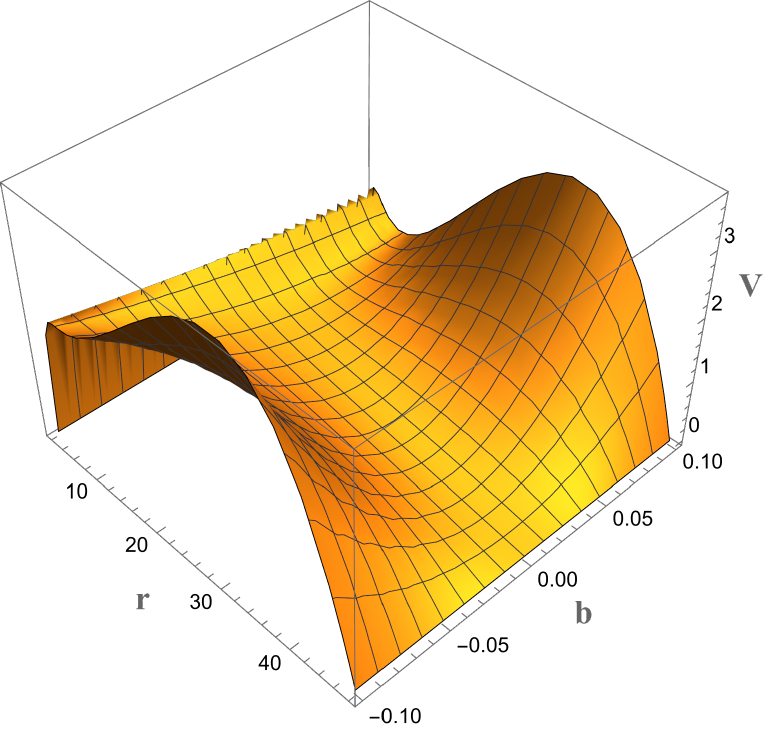}
    \end{subfigure}
    \hfill
    \begin{subfigure}{0.49\textwidth}
        \centering
        \includegraphics[scale=0.5, trim=0cm -2cm 0cm -3cm, clip]{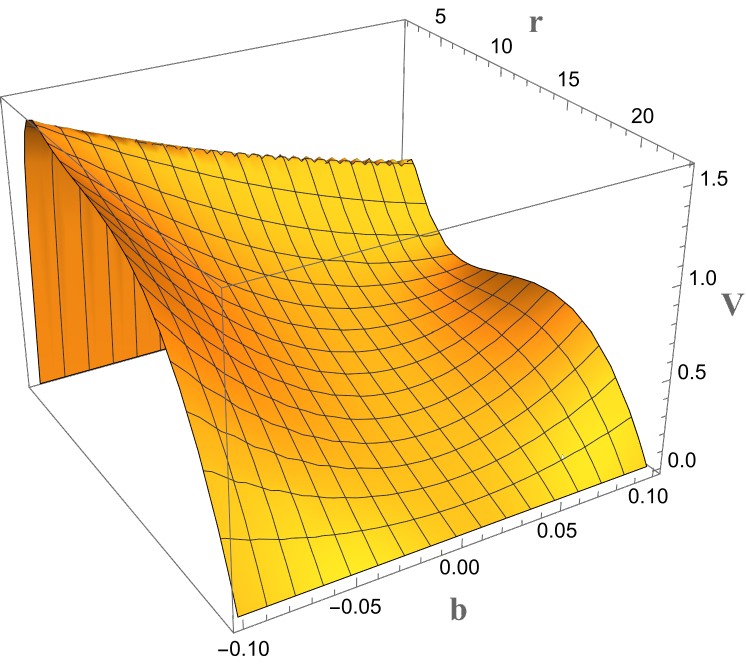}
    \end{subfigure}
    \caption{The effective potential (\ref{V0}) as a function of $b$ and $r \in [r_- , r_+ ]$. The value of the quintessence parameter is $k=0.015$ in the left panel and $k=0.04$ in the right panel. The other numerical values are: $M=1$, $w=-2/3$ and $L=6$. }
    \label{fig:pot3D}
\end{figure}

The energetic boundary $E^2=V$ forms a closed curve for the trapped charged particle whose trajectory is restricted to a region governed by the effective potential (\ref{V}), which in the equatorial plane is given in (\ref{V0}).
As an example, let us represent in the figure \ref{fig:pot3D} the potential $V_0$ defined in (\ref{V0}).
As it can be noticed, the shape of the potential is strongly depending on the model's parameters and one has to find ranges for $b$ and $k$ 
for which the particle can be trapped on bound orbits. For small values of the quintessence parameter $k$, there is a second maximum of the potential, close to the cosmological horizon which is most prominent for large values of $|b|$ (see the left panel of the figure \ref{fig:pot3D}). The existence of a minimum value of the potential between two maxima allows the particle with suitable energy to move on stable bound orbits. If $k$ is increasing, the second maximum vanishes (see the right panel of the figure \ref{fig:pot3D}). No stable circular orbits exist and, depending on the starting point, the particle's trajectory can be either a capture or an escape one.

Once the model's parameters $M$, $b$, $w$ and $k$ are fixed, one may have different types of trajectories, depending on the particle's energy, specific charge and angular momentum.  A detailed analysis of the potential and classification of orbits of charged particles for the general line element (\ref{metmag}) has been developed in \cite{Lungu:2024iob}. The motion of photons and the classification of orbits in the spacetime described by the metric (\ref{metmag}) was performed in \cite{Lungu:2025iri}.

\subsection{Circular orbits in the equatorial plane}

In the followings, let us discuss the circular orbits in the equatorial plane by working out the relation $V_0^{\prime} (r=r_c ) =0$, where $V_0$ is defined in (\ref{V0}). If $V_0^{\prime \prime} (r=r_c)>0$, the corresponding orbit is stable, otherwise it is unstable. For $\theta = \pi/2$, the relation (\ref{Lr}) leads to the followings quadratic equation with respect to the
angular momentum
\begin{eqnarray}
(f'r-2f) L^2-2 b  r^3 f' L
+r^3\left[( b^2 r^2+1) f' +2 b^2 fr \right]=0
\label{Lr0}
\end{eqnarray}
The solutions correspond to the angular momentum of the charged particle moving on a circular orbit of radius $r=r_c$
\begin{equation}
L_{\pm}=\frac{r \left(f^{\prime}r^2 b  \pm \sqrt{(4f^2 b^2 -f'{^2})r^2+2ff^{\prime} r}\right)}{f^{\prime}r-2f}
\label{Lpm}
\end{equation}
with the energy
\begin{equation}
E_{\pm}^2=\frac{2f^2}{\left(f^{\prime}r-2f\right)^2} 
\left[2f(1+2 b^2  r^2 ) -r f^{\prime} \pm 2 r b \sqrt{(4f^2 b^2 -f'{^2})r^2+2ff^{\prime} r}\right] 
\label{Epm}
\end{equation}
Obviously, one has to impose that $L_{\pm}$ are real and this happens for $(4f^2 b^2 -f'{^2})r^2+2ff^{\prime} r \geq0$. For a general metric function of the form (\ref{fKis}), one obtains a constrain on the magnetic parameter for the existence of circular orbits:
\begin{equation}
b^2  \geq \frac{f^{\prime} (r f^{\prime}-2f)}{4f^2 r} .
\label{B2}
\end{equation}
Once this relation is satisfied, the energies $E_{\pm}^2$ defined in (\ref{Epm}) are real and positive quantities.
For a giving value of $w$, the condition (\ref{B2}) leads to ranges for the circular radius.

As a first particular case, let us take $w=-2/3$ for which
the condition (\ref{B2}) becomes
\begin{equation}
b^2 \geq  \frac{(2M-kr^2)(6M-2r+kr^2)}{4r^2(2M-r+kr^2)^2} 
\label{x}
\end{equation}
One may notice that the right hand side of the above expression is vanishing in $r=r_*= \sqrt{2M/k}$, with $r_*$ being static radius defined in (\ref{rmax}), and also in
\begin{equation}
r_1 = \frac{1- \sqrt{1-6kM}}{k} \, , \quad r_2 = \frac{1+ \sqrt{1-6kM}}{k}
\end{equation}
One has the relation
\[
r_- < r_1 < r_* < r_+ < r_2
\]
For $r \in ( r_1 , r_* ) $, the right hand side of (\ref{x}) is negative and therefore there are no constrains on the magnetic parameter $b$. Indeed, for small values of $r$, the potential always allows an unstable circular orbit corresponding to its maximum value located close to $r_-$. On the other hand, for $r \in ( r_* , r_+ )$, the right hand side of (\ref{x}) becomes positive and a magnetic field is needed, satisfying the relation (\ref{x}), in order to compensate the effect of quintessence. In view of the condition (\ref{rmax}), this orbit is also unstable and it corresponds to the maximum of the potential located close to $r_+$. If the potential has two maxima, the minimum between them allows the existence of a stable circular orbit with $r<r_*$.

The second example corresponds to $w=-1$ for which $r_*$ given in (\ref{rmax}) is $r_* = \left( \frac{M}{k} \right)^{1/3}$.
The condition (\ref{B2}) becomes
\begin{equation}
b^2 \geq \frac{(3M-r)(M-kr^3)}{r^2(2M-r+kr^3)^2} 
\label{y}
\end{equation}
and, for $3M < r < r_*$, the right hand side of (\ref{y}) is negative and there are no constrains on the magnetic parameter.
For any other range of the radius for which $f>0$, a magnetic field satisfying (\ref{y}) should be taken into account.

On the other hand, the physical condition for a stable circular orbit in the equatorial plane, $\partial_{\theta}^2 V|_{\theta=\pi/2}>0$, leads to the important constrain 
\begin{equation}
L >L_0= |b| r^2
\label{L0r}
\end{equation}
From the above relation together with (\ref{Lrange}), one obtains the range of the stable circular obit angular momentum
\begin{equation}
L_0<L<L_{1,2}^+.
\label{Lrangecirc}
\end{equation}

\subsection{The innermost stable circular orbit and parameters ranges}

Because of its astrophysical relevance, let us discuss the innermost stable circular orbit (ISCO) which is the smallest marginally stable circular orbit followed by a test particle orbiting the black hole. Its location is depending on the model's parameters being given by the inflection point of the effective potential, $V_0^{\prime}=0$ and $V_0^{\prime \prime}=0$. Thus, we obtain a system made of the equation (\ref{Lr0}) with the solution (\ref{Lpm}) and the equation
\begin{equation}
\left(f''r^2-4f'r+6f\right)L_1^2-2bf''r^4L_1+\left[\left(f''r^2+4f'r+2f\right)b^2+f''\right]r^4=0
\label{Vsec}
\end{equation}
with the solution
\begin{equation}
L_{1\pm}=\frac{bf''r^4 \pm\sqrt{4b^2\left(-2f''fr^2+4f'{^2}r^2+4f'fr+3f^2\right)-f''\left(f''r^2-4f'r+6f\right)}}{f''r^2-4f'r+6f}
\label{L1pm}
\end{equation}
By using Cramer's rule, we find the common solution of these two equations,  which corresponds to $L_{ISCO}$, as being:
\begin{equation}
L_{ISCO}=\frac{1+2b^2r^2}{2b}+\frac{2bfr\left(f-f'r\right)}{f''fr-2f'{^2}r+3ff'}
\label{Lisco}
\end{equation}
with the corresponding radius $r=r_{ISCO}$.

\begin{figure}[H]
    \centering
    \begin{subfigure}{0.49\textwidth}
        \centering
        \includegraphics[scale=0.5, trim=2cm 12cm 5cm 1cm, clip]{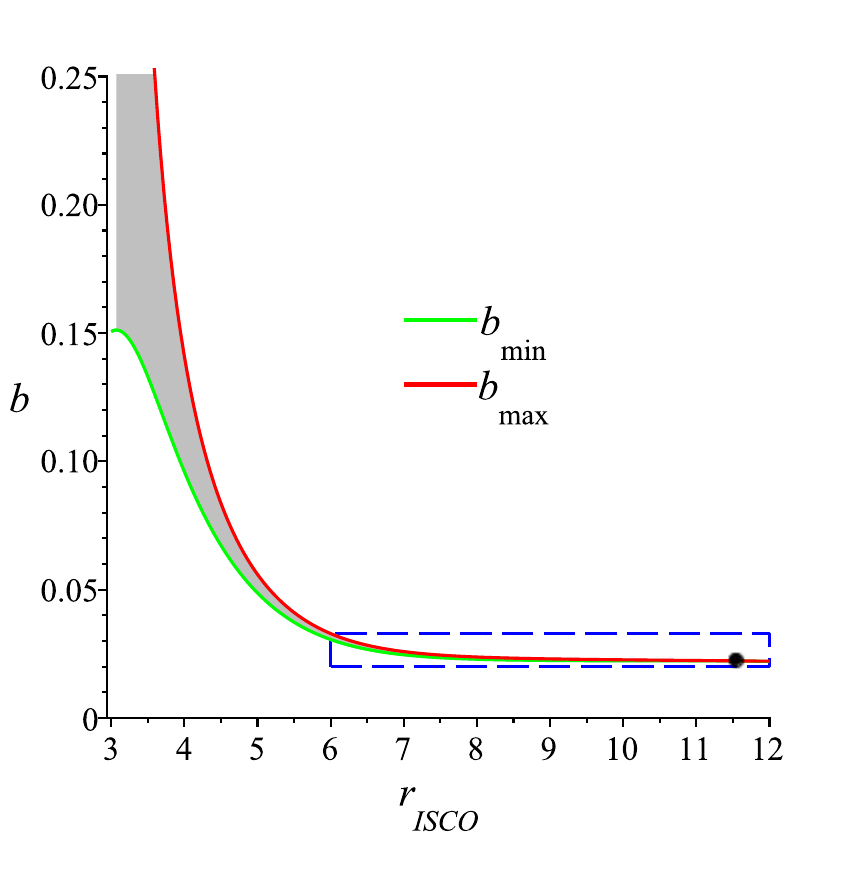}
    \end{subfigure}
    \hfill
    \begin{subfigure}{0.49\textwidth}
        \centering
        \includegraphics[scale=0.5, trim=0cm 12cm 5cm 1cm, clip]{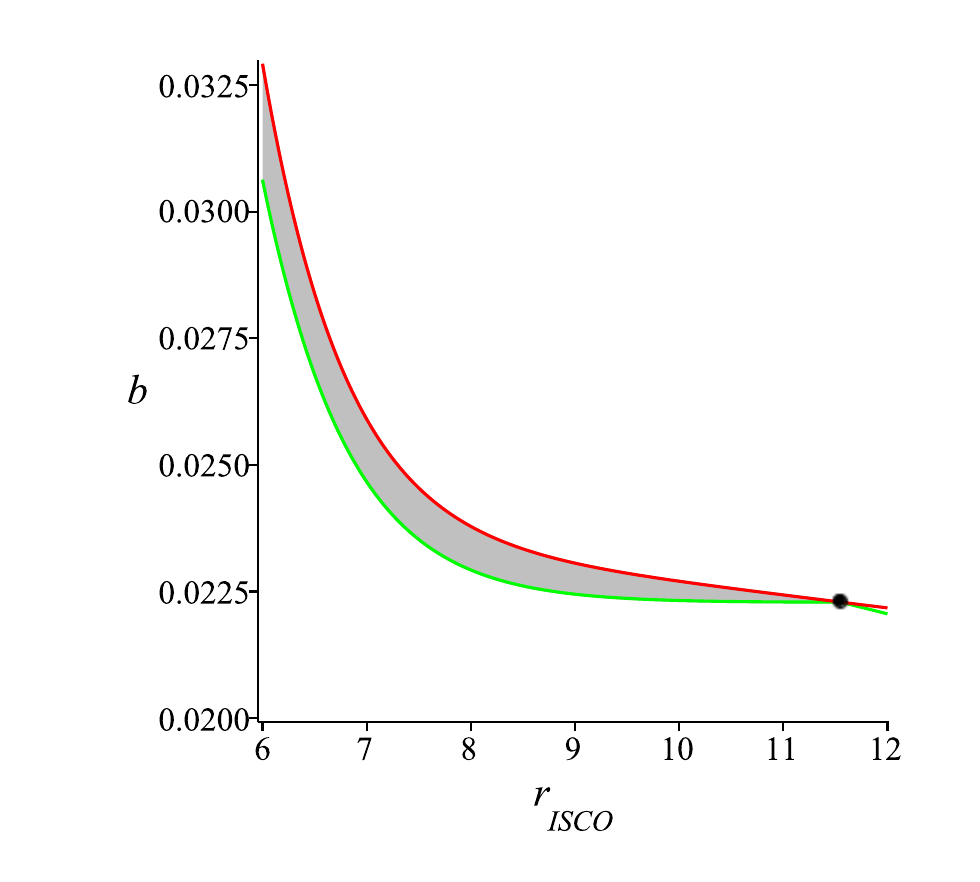}
    \end{subfigure}
    \begin{tikzpicture}[overlay]
\node at (5, 4){};
\draw[->, thick] (-7,5.1) -- (-12, 2.2);
\end{tikzpicture}
    \caption{Representation of the allowed range of $b$ (greyed area) as a function of the radius $r_{ISCO}$. The black dot represents the radial coordinate $r_*=11.55$ for which $b_{min}=b_{max}$. The values of the parameters are: $M=1$, $w=-2/3$ and $k=0.015$.}
    \label{fig:breg}
\end{figure}

Let us consider a positive $b$, but the similar results can be obtained for $b<0$.  By imposing the stable circular orbit condition (\ref{Lrangecirc}), i.e. $L_0<L_{ISCO}<L_{1}^{+}$, one finds that the magnetic parameter $b$ should lie in the range 
\begin{equation}
b_{min}<b<b_{max},
\label{brange}
\end{equation}
where
\begin{equation}
b_{min}=\frac{\sqrt{f_s/f-1} \left(f''fr-2f'{^2}r+3f'f\right)r}{4fr(f'r-f)}\left[\sqrt{1+\frac{4fr\left(f'r-2f\right)}{\left(f_s/f-1\right)\left(f''fr-2f'{^2}r+3f'f\right)r^2}}-1\right]
\label{bmin}
\end{equation}
and
\begin{equation}
b_{max}=\sqrt{\frac{f''fr-2f'{^2}r+3f'f}{4fr\left(f'r-2f\right)}}.
\label{bmax}
\end{equation}
An example is given in the figure \ref{fig:breg}. As it can  be noticed, at the reference value $r=r_*$, one has $b_{min} = b_{max}$.

The key results that determine the range of bound orbits are summarized in table \ref{table1}, where $L_0= |b| r^2$ and $L_{1,2}^+$ are defined in (\ref{L1}) and (\ref{L2}). The conditions for $r_{ISCO}$ and $k$ are obtained by imposing that $b_{min}$ and $b_{max}$ are real and positive quantities and these are given in table \ref{tab:isco}. 
\begin{table}[h!]
\centering
\begin{tabular}{|c|c|c|}
\hline
 & $b>0$ & $b<0$ \\ \hline
$L$ & $L_1^-<L<L_1^{+}$ & $L_2^-<L<L_2^{+}$ \\ \hline
$r$ & \multicolumn{2}{|c|}{$r_{ISCO}<r<r_*$} \\ \hline
\end{tabular}
\caption{Constrains on $b$, $L$ and $r$ required for the existence of bound orbits.}
\label{table1}
\end{table}

\begin{table}[h!]
\renewcommand{\arraystretch}{1.6}
\centering
\begin{tabular}{|c|c|c|}
\hline
$w$ & $r_{ISCO}$ & $k$ \\ \hline
\multirow{3}{*}{$-2/3$} & $(3M,4M)$ & $0<k<\frac{2(r-3M)}{r^2}$ \\ \cline{2-3}
                                         & $[4M,6M)$ & $0<k< \frac{2M}{r^2}$ \\ \cline{2-3}
                                         & $[6M,r_*)$ & $ \frac{1}{2r^2}\left[3(r-4M)-\sqrt{9r^2-80Mr+192M^2}\right]  <k<\frac{2M}{r^2}$\\ \hline
\multirow{2}{*}{$-1$} & $(3M,6M)$ & $0<k<\frac{M}{r^3}$ \\ \cline{2-3}
                                         & $[6M,r_*)$ & $ \frac{M(r-6M)}{r^3(4r-15M)}<k<\frac{M}{r^3}$ \\ \hline
                                         
\end{tabular}
\caption{$r_{ISCO}$ ranges and constrains on the parameter $k$, for $b>0$ defined in the range (\ref{brange}).}
\label{tab:isco}
\end{table}

\begin{figure}[H]
    \centering
    \begin{subfigure}{0.49\textwidth}
        \centering
        \includegraphics[scale=0.5, trim=2cm 12cm 5cm 1cm, clip]{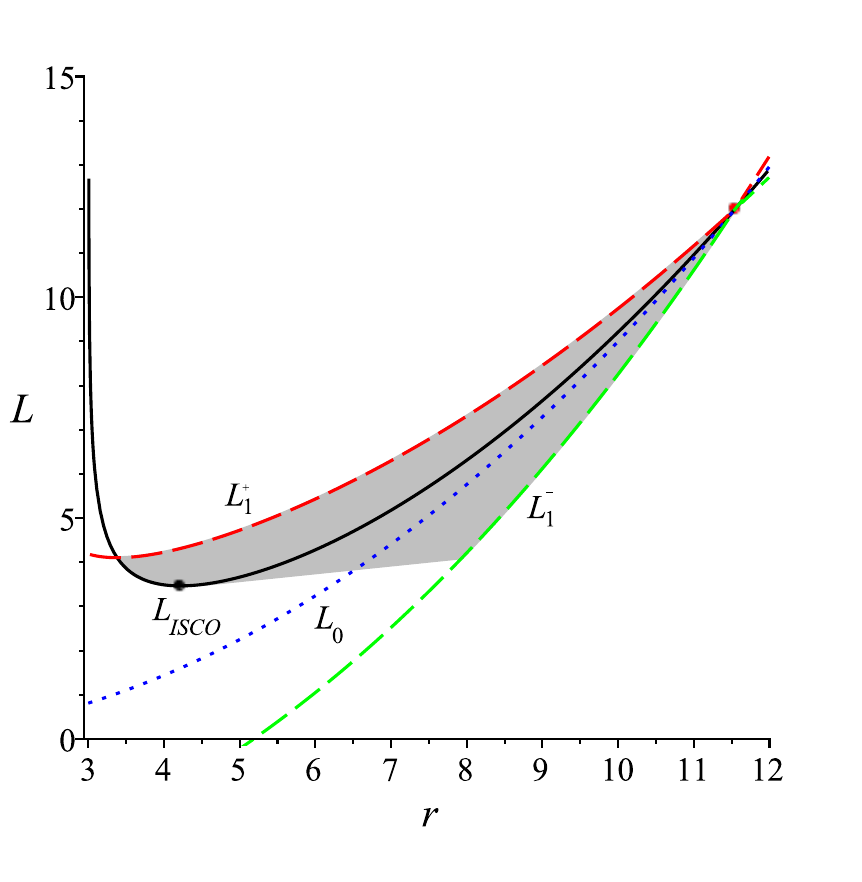}
    \end{subfigure}
    \hfill
    \begin{subfigure}{0.49\textwidth}
        \centering
        \includegraphics[scale=0.5, trim=0cm 12cm 5cm 1cm, clip]{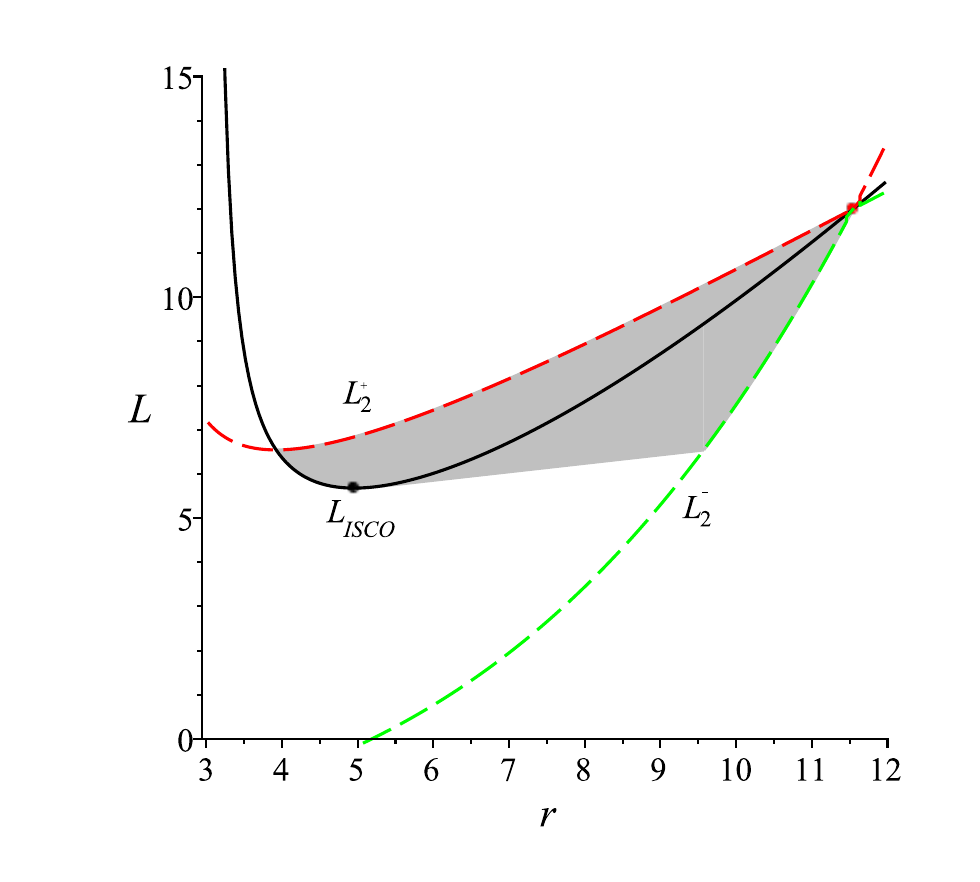}
    \end{subfigure}
    \caption{ Representation of trapped states (shaded area) corresponding to the condition $L_{1,2}^{-}<L<L_{1,2}^{+}$. The blue dotted line represents $L_0$, corresponding to $r_0$, where $\dot{\phi}$ changes sign. The solid black line is the angular momentum $L_+$ defined in (\ref{Lpm}), for $b=0.09$ (left panel) and $b=-0.09$ (right panel). The numerical values are: $M=1$, $w=-2/3$ and $k=0.015$.}
    \label{fig:Lreg}
\end{figure}

An example is given in the figure \ref{fig:Lreg} where the shaded area is representing the allowed region of trapped states. Circular orbits lie within this shaded region, and the minimum of $L_+$ defined in (\ref{Lpm}) corresponds to ISCO. It can be seen that bound orbits are not allowed for $L$ below $L_{ISCO}$, which represents the minimum angular momentum required for a charged particle to remain in a stable circular orbit around the black hole. Conversely, as shown in the right panel of figure \ref{fig:Lreg}, for $b<0$, larger values of the angular momentum are required to obtain trapped orbits, compared to the case $b>0$ (see the left panel of figure \ref{fig:Lreg}). In both cases, there is a maximum value of $L=L_{max}$, corresponding to $r=r_*$ for which bound orbits are allowed. The blue dotted line separates the region where $\dot{\phi}>0$ (above) from the one where $\dot{\phi}<0$ (below). 

\begin{figure}[H]
    \centering
    \begin{subfigure}{0.49\textwidth}
        \centering
        \includegraphics[scale=0.5, trim=2cm 12cm 5cm 1cm, clip]{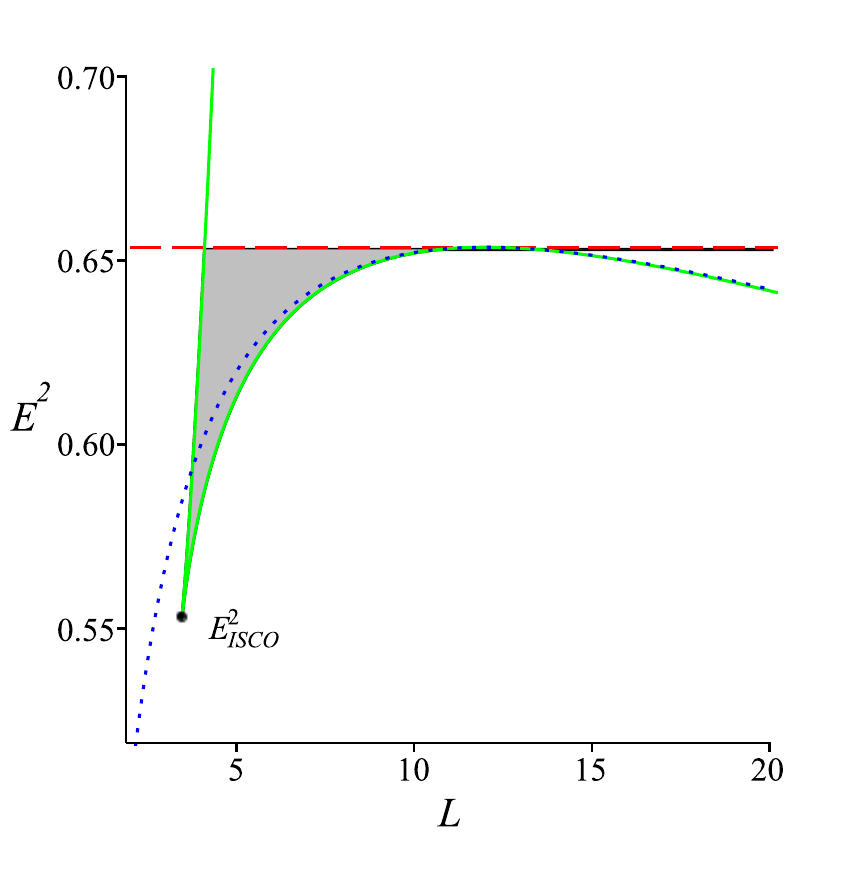}
    \end{subfigure}
    \hfill
    \begin{subfigure}{0.49\textwidth}
        \centering
        \includegraphics[scale=0.5, trim=0cm 12cm 5cm 1cm, clip]{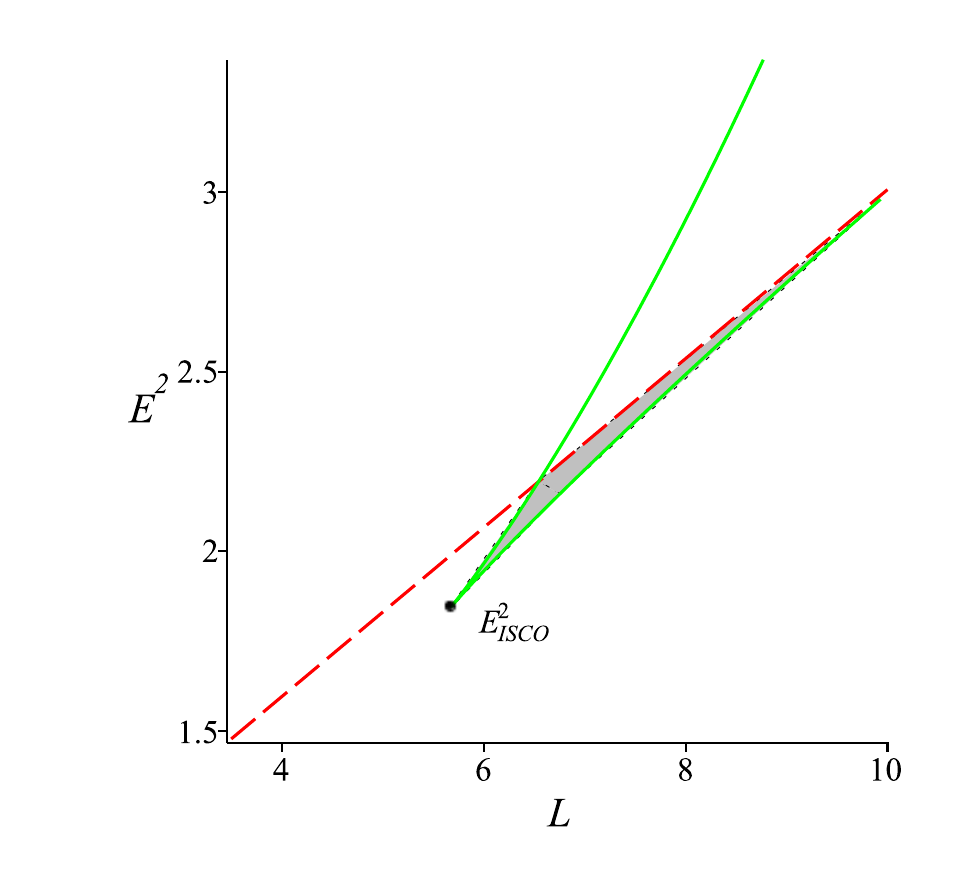}
    \end{subfigure}
    \caption{Energetics of bound orbits for $b=0.09$ (left panel) and $b=-0.09$ (right panel). The pairs $\lbrace E^2, L\rbrace$ corresponding to bound orbits are situated in the shaded region. The green curve corresponds to circular orbits.  The red dashed line represents the energy corresponding to the saddle points (\ref{Es}) and (\ref{Es1}) respectively. The blue dotted line corresponds to $E_0$ defined in (\ref{E0}) and represents the boundary between curly (for $E^2>E_0^2$) and non curly (for $E^2<E_0^2$) trajectories. The numerical values are: $M=1$, $w=-2/3$ and $k=0.015$.}
    \label{fig:Ereg}
\end{figure}

In figure \ref{fig:Ereg}, the allowed region of motion is shown in the $E^2$–$L$ parameter space. The minimum energy corresponds to $E_{{ISCO}}^2$ and bound orbits exist for $E^2 < E_{s1,s2}^2$, where $E_{s1,s2}^2$ are indicated by the dashed lines. In the case of $b>0$ (see the left panel), bound orbits occur for lower values of the energy $E^2$, compared to the case $b<0$ (see the right panel). The minimum energy $E_0$ required for a charged particle to exhibit a curly trajectory is defined in (\ref{E0}). As shown in the figure \ref{fig:Ereg}, trajectories near the ISCO are non-curly, whereas those farther away become curly.

\begin{figure}[H]
    \centering
    \begin{subfigure}{0.49\textwidth}
        \centering
        \includegraphics[scale=0.5, trim=2cm 12cm 5cm 1cm, clip]{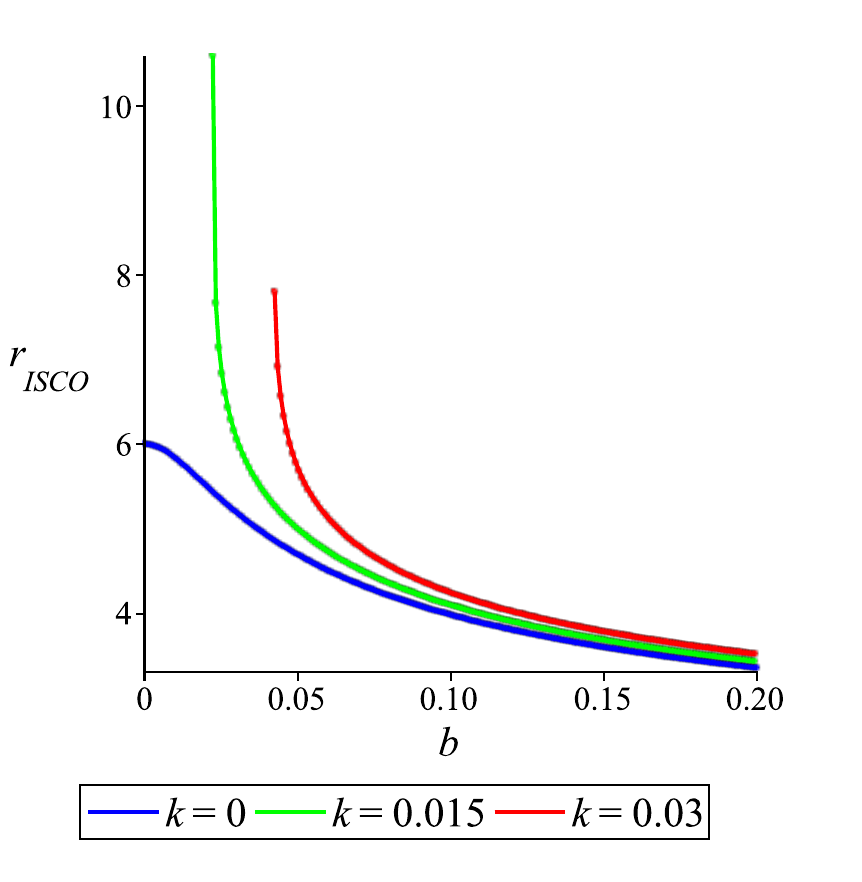}
    \end{subfigure}
    \hfill
    \begin{subfigure}{0.49\textwidth}
        \centering
        \includegraphics[scale=0.5, trim=2cm 12cm 5cm 1cm, clip]{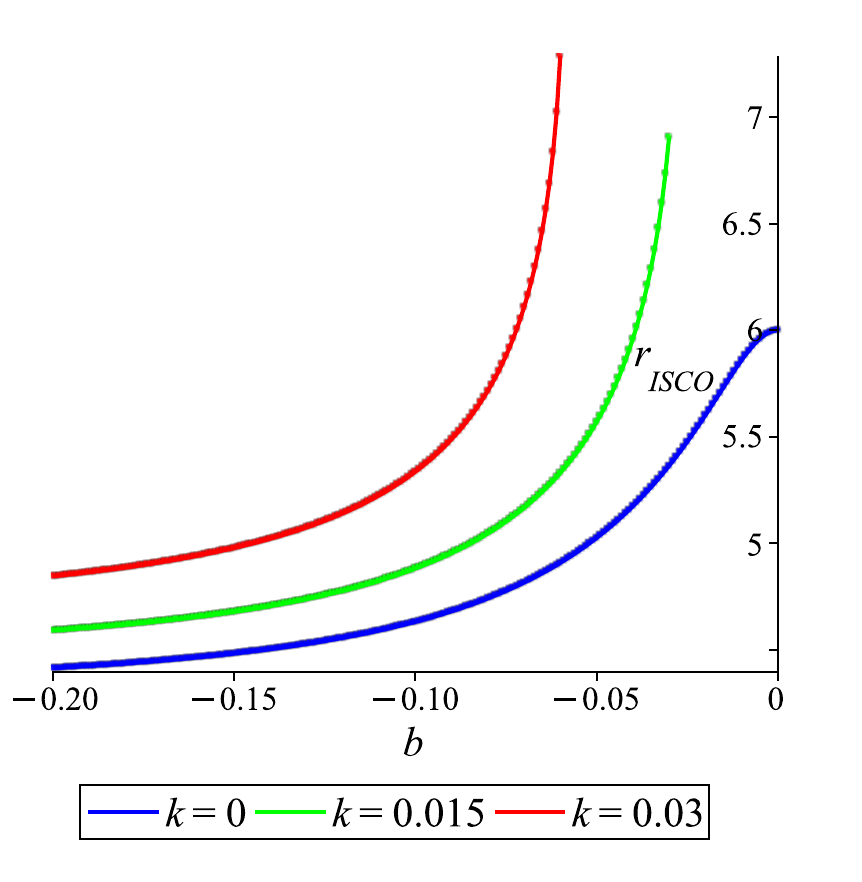}
    \end{subfigure}
    \caption{Position of $ISCO$ in dependence on the magnetic parameter $b$, for different values of $k$, for $b>0$ (left panel) and $b<0$ (right panel). The other numerical values are $M=1$ and $w=-2/3$.}
    \label{fig:riscob}
\end{figure}

In view of the analysis developed in this section, one may conclude by saying that the stable circular trajectories, if exist, are in the region $r \in [r_{ISCO} , r_*]$, where $r_*$ is depending only on the quintessence parameters $w$ and $k$ (see the formula (\ref{rmax})). In what it concerns the radius $r_{ISCO}$, this is strongly depending on both parameters $b$ and $k$ which have to be in their allowed ranges given in (\ref{brange}) and table \ref{tab:isco}. An illustrative example is shown in figure \ref{fig:riscob}. For $k=0$ and $b=0$, one obtains the well known result $r_{ISCO}=6M$. For the Ernst spacetime (the blue plots), all values of $r_{ISCO}$ are located below $r=6M$ \cite{Kolos:2015iva}. However, in the presence of quintessence, values of $r_{ISCO}$ bigger than $6M$ are possible, the maximum value of $r_{ISCO}$ being $r_*$ given in (\ref{rmax}).  For a fixed value of $b$, $r_{ISCO}$ is increasing with $k$, while for a given value of $k$, as $|b|$ increases, $r_{ISCO}$ decreases.

\section{Quasi-harmonic oscillations}

\subsection{Fundamental frequencies}

In this section, let us discuss the important problem of 
the epicyclic motion of the charged particle
governed by linear harmonic oscillations around the minimum of the effective potential defined in (\ref{V}).
This issue has deep implications in correlating the theory with observational data.
Even though the stable quasi-circular orbits have a regular (harmonic)
character, as it has been pointed out for the Ernst spacetime  \cite{Frolov:2010mi, Kolos:2015iva}, the effect of quintessential matter has a deep impact on the shape and stability of orbits \cite{Lungu:2024iob}.

Thus, let us consider the charged particle moving on a stable circular orbit in the equatorial plane, whose angular momentum and energy are given by (\ref{Lpm}) and (\ref{Epm}).  When it is slightly perturbed from its stable circular orbit, the particle starts to oscillate around the circular orbit of radius $r_c$ and with respect to the equatorial plane, i.e. $r = r_c + \delta r$ and $\theta = \frac{\pi}{2} + \delta \theta $. The small perturbations $\delta r$ and $\delta \theta$ are satisfying the equations of a linear harmonic
oscillator, i.e.
\begin{equation}
\delta \ddot{r} + \omega_r^2 \delta r = 0 \;  , \quad \delta \ddot{ \theta } + \omega_{\theta}^2 \delta \theta = 0 
\label{rt}
\end{equation}
The dot in the above
equation denotes the derivatives with respect to the proper
time and these frequencies characterizing the epicyclic oscillations are measured by a local observer.

Let us mention that, using the first-order perturbative approach developed in \cite{Lungu:2024iob}, one may consider the first equation in (\ref{rt}) as the one describing a simple harmonic oscillator motion. Its solution is inducing variations of the frequency $\omega_{\theta}^2$ and the latitudinal motion is described by a Mathieu's type equation. This motion is resonantly excited by the radial oscillations and the parametric resonances appear for specific values of the ration $\omega_r / \omega_{\theta}$. As it has been proved in \cite{Lungu:2024iob}, for $r< r_* = \sqrt{2M/k}$, there are large regions of stability. 
The first and
strongest resonance occurs for $\omega_r / \omega_{\theta}=2/3$ and this case was widely discussed in literature (see for example \cite{Abramowicz:2002xc}).

In the followings, let us compute the fundamental frequencies 
of test particles moving on quasi-circular orbits around the magnetized black hole immersed in quintessence and analyze the effects of the model's parameters 
on them.
Since our background is stationary spherically symmetric, the radial and
latitudinal frequencies measured by a local observer are
defined by the relations:
\begin{equation}
\omega_r^2 = \frac{1}{2} \frac{\partial^2 V}{\partial r^2}  
\label{omegar}
\end{equation}
and
\begin{equation}
\omega_{\theta}^2 = \frac{1}{2 r^2 f} \frac{\partial^2 V}{\partial \theta^2}  
\label{omegat}
\end{equation}
while the angular frequency can be derived from (\ref{Lw}) as being
\begin{equation}
\omega_{\phi} = \dot{\phi} = \frac{L}{r^2} - b 
\label{omegaf}
\end{equation}
In physical units, the dimensionless form
of the above frequencies reads
\[
\nu = \frac{1}{2 \pi} \frac{c^3}{GM} \omega
\]
For a general metric function (\ref{fKis}), the expressions (\ref{omegar}) and (\ref{omegat}) computed in the equatorial plane are:
\begin{equation}
\omega_r^2 =  \frac{1}{2} \left \lbrace b^2  \left[ r^2 f^{\prime \prime} + 4r f^{\prime} +2f \right]
-2 b  f^{\prime \prime} L + \frac{L^2}{r^4} \left[ r^2 f^{\prime \prime} -4r f^{\prime} + 6f \right] + f^{\prime \prime}
\right \rbrace
\label{omr}
\end{equation}
and
\begin{equation}
\omega_{\theta}^2 =  \frac{L^2}{r^4} -b^2
\label{omt}
\end{equation}
For the angular momentum satisfying the constrain (\ref{Lrangecirc}), the quantity $\omega_{\theta}^2$ is positive and one has stable latitudinal oscillations. In the followings, we shall consider the positive root of $L$ defined in (\ref{Lpm}), while $b$ may have either positive or negative values. As we mentioned previously, the other cases corresponding to $L<0$ are equivalent to these ones.

In order to relate the locally measured frequencies defined in (\ref{omegaf}), (\ref{omr}) and (\ref{omt}) to the ones measured by a distant observer one has to employ the
gravitational redshift transformation 
\begin{equation}
\Omega_{r , \theta , \phi} = \frac{\omega_{r , \theta , \phi}}{\dot{t}} = \frac{f}{E} \omega_{r , \theta , \phi}
\end{equation}
where $f$ and $E$ are respectively defined in (\ref{fKis}) and (\ref{Epm}), with $r$ being the radius of the circular orbit.
The above frequencies measured by observers located at infinity can be compared to observational data.
As discussed in the previous section, the radial and latitudinal frequencies are defined on the region of existence of the stable circular orbits, being limited by ISCO
where the radial frequency vanishes and by $r_*$ where $\Omega_{\theta}$ vanishes.
Using the expression (\ref{omt}) with $f$ given in (\ref{metric}), we obtain
the simple relation
\begin{equation}
\Omega_{\theta}^2 = \frac{M}{r^3} - \frac{k}{2r}
\end{equation}
One may notice that the latitudinal frequency does not depend on $b$ and this can be understood by the fact that the Lorentz force is acting only on the radial direction. However, in the presence of quintessence, $\Omega_{\theta}^2$ decreases with $k$. 

At $r=r_* = \sqrt{2M/k}$, the quantity $\Omega_{\theta}$ is vanishing. Thus, one may conclude by saying that the static radius $r=r_*$ is separating the region $r<r_*$ where $\Omega_{\theta}^2>0$ and the circular trajectories are stable under fluctuations in the transverse direction from the one corresponding to $r>r_*$ where $\Omega_{\theta}^2<0$. In this case, there are no vertically stable orbits and the
particle is accelerated along the $Oz-$axis toward infinity (see the equations (\ref{rt})).

On the other hand, the radial frequency $\Omega_r$ is characterizing the oscillatory motion in the
equatorial plane under the influence of the Lorentz force, attracting gravitation and repulsive quintessence. Thus, $\Omega_r^2$
has a more complicated form, being dependent on both $b$ and $k$. For $b=0$ and $k=0$, the expressions corresponding to
the Schwarzschild case are recovered, i.e.
\[
\Omega_r^2 = \frac{M(r-6M)}{r^4} \, , \quad \Omega_{\theta}^2 = \frac{M}{r^3}
\]

\subsubsection{The positive magnetic parameter}

\begin{figure}[H]
\centering
\includegraphics[width=0.4\textwidth]{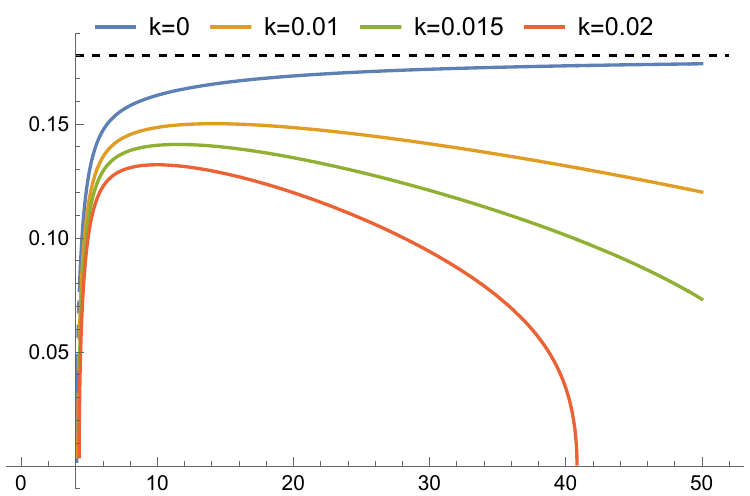}
\includegraphics[width=0.4\textwidth]{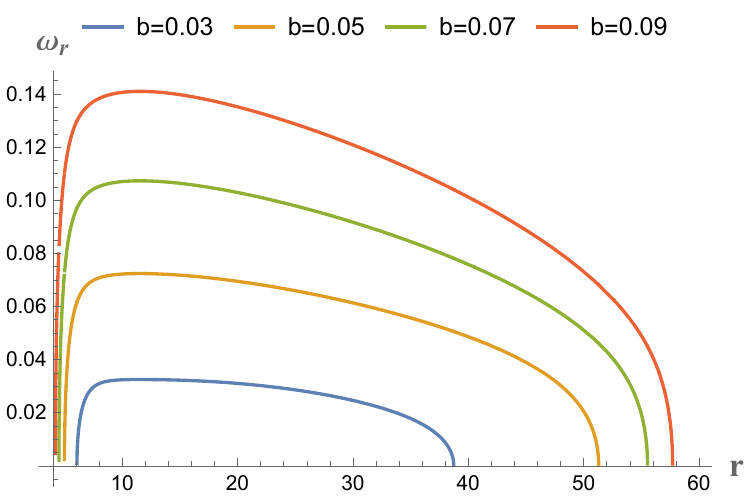}
\caption{{\it Left panel}. The radial profile of $\omega_r$ for $b=0.09$ and different values of $k$. {\it Right panel}. The radial profile of $\omega_r$ as a function of $r$, for $k=0.015$ and different values of $b$. The other numerical values are: $M=1$ and $w=-2/3$.}
\label{omR}
\end{figure}

Firstly, let us consider the case corresponding to $b>0$ for which the Lorentz force acting on the charged particle is repulsive being directed outward from the
black hole.
In the left panel of the figure \ref{omR} one may notice that, contrary to Ernst spacetime where $\omega_r$ is increasing with $r$ and tends to the Larmor frequency $\omega_r \to \omega_L = 2b$ for large values of $r$ (the dashed horizontal line), once the quintessence comes into place, there is a range of $r$ for which $\omega_r^2$ given in (\ref{omr}) is positive. As $k$ increases, the values of $\omega_r$ get smaller and the range of $r$ allowing stable radial oscillations is shrinking (see the left panel). There is a critical value of $k$, depending on $b$, above which the condition $\omega^2_r > 0$ can not be satisfied and the system becomes unstable in the radial direction. 
In what it concerns the dependence on $b$, one may notice in the right panel of the figure \ref{omR} the opposite behaviour, in the sense that the values of $\omega_r$ and its range are increasing with increasing $b$.  
There is a critical value of $b$, depending on $k$, below which the condition $\omega^2_r > 0$ can not be satisfied. However, the conditions given in table \ref{table1} should be verified.

\begin{figure}[H]
\centering
\includegraphics[width=0.45\textwidth]{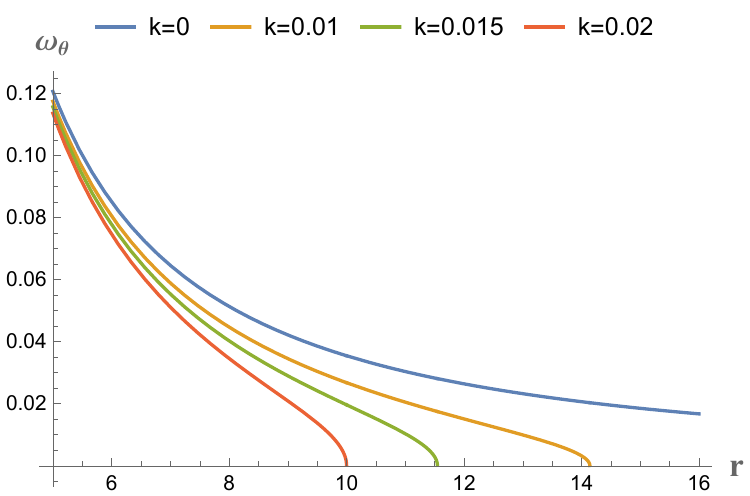}
\includegraphics[width=0.45\textwidth]{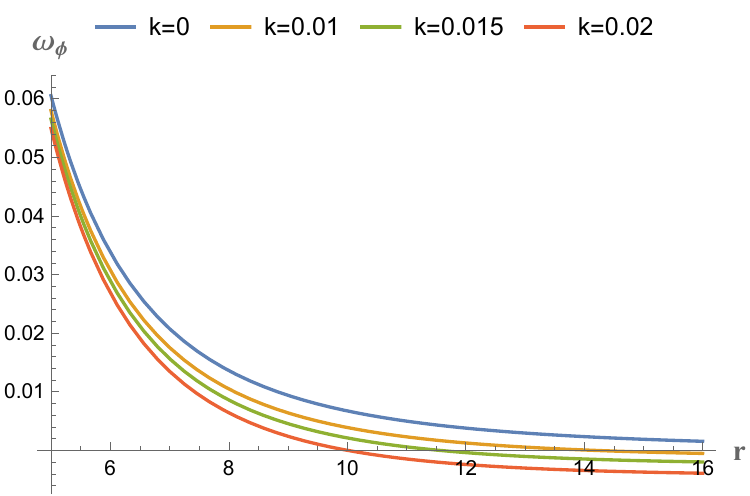}
\caption{{\it Left panel}. The expressions of $\omega_{\theta}$ (the left panel) and $\omega_{\phi}$ (the right panel) as functions of $r$, for different values of $k$. The other numerical values are: $M=1$, $b=0.09$ and $w=-2/3$.}
\label{omtf}
\end{figure}

\begin{figure}[H]
\centering
\includegraphics[width=0.45\textwidth]{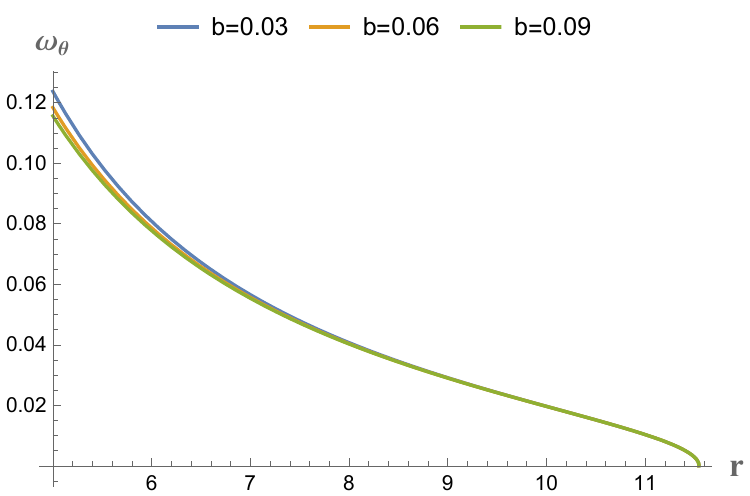}
\includegraphics[width=0.45\textwidth]{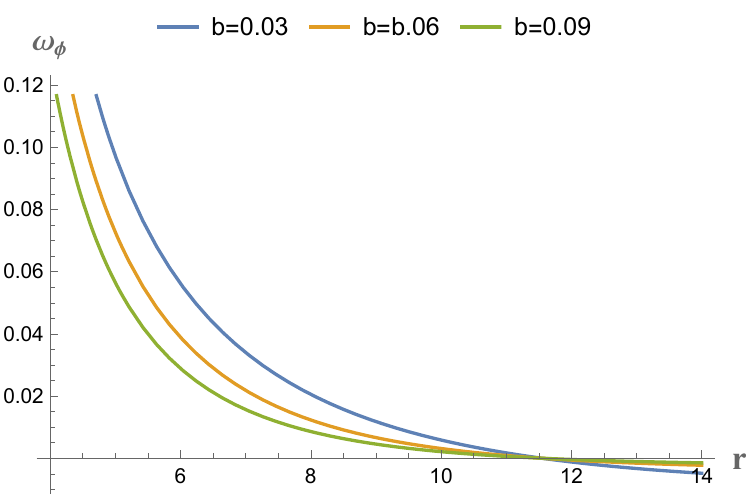}
\caption{{\it Left panel}. The expressions of $\omega_{\theta}$ (the left panel) and $\omega_{\phi}$ (the right panel) as functions of $r$, for different values of $b$. The other numerical values are: $M=1$, $k=0.015$ and $w=-2/3$.}
\label{omtfb}
\end{figure}

Even though the expressions of $\omega_{\theta}^2$ and $\omega_{\phi}$ are same same as for the Ernst spacetime \cite{Kolos:2015iva}, the contribution of quintessence is encoded in the angular momentum (\ref{Lpm}). As it can be noticed in the figures \ref{omtf} and \ref{omtfb}, both $\omega_{\theta}$ and $\omega_{\phi}$ are monotonically decreasing with $r$ and this behavior is more prominent as $k$ or $b$ increases. For $k \neq 0$, there is a critical value of $r$, denoted by $r_*$ and defined in (\ref{rmax}), where $\omega_{\theta}$ vanishes and the frequency $\omega_{\phi}$ changes the sign. However, the value $r_*$ is inside the range where $\omega_r$ is defined and, in view of the analysis developed in the previous section, the allowed range of $r$ allowing the existence of stable circular orbits is $r \in [ r_{ISCO} , r_*]$. In what it concerns $r_{ISCO}$ where $\omega_r$ is vanishing, the conditions given in table \ref{tab:isco} should be satisfied.

\subsubsection{The negative magnetic parameter}

\begin{figure}[H]
\centering
\includegraphics[width=0.45\textwidth]{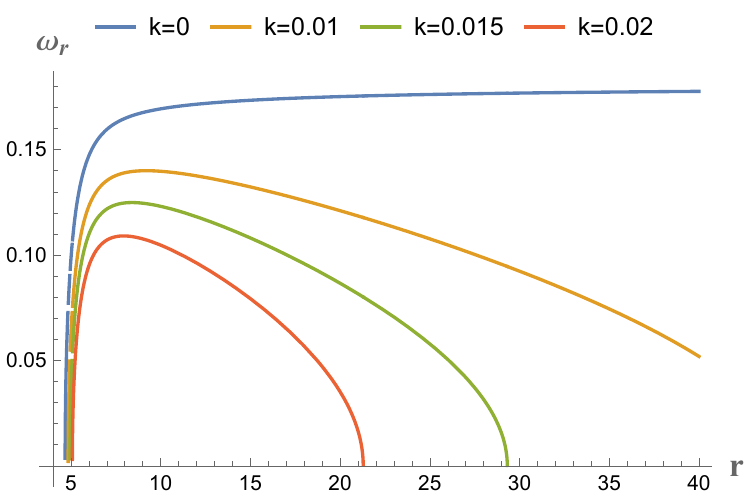}
\includegraphics[width=0.45\textwidth]{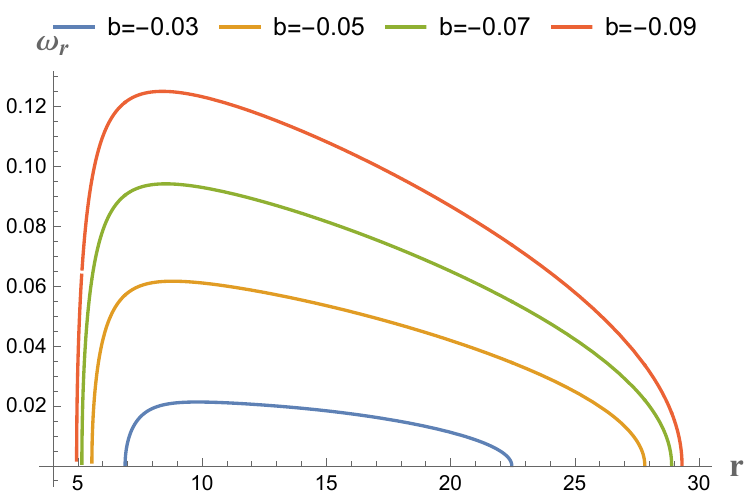}
\caption{The frequency $\omega_r$ as a function of $r$, for different values of $k$ and $b=-0.09$ ({\it Left  panel}) and for different values of $b$ and $k=0.015$ ({\it Right panel}). The other numerical values are: $M=1$  and $w=-2/3$.}
\label{omrbm}
\end{figure}

\begin{figure}[H]
\centering
\includegraphics[width=0.45\textwidth]{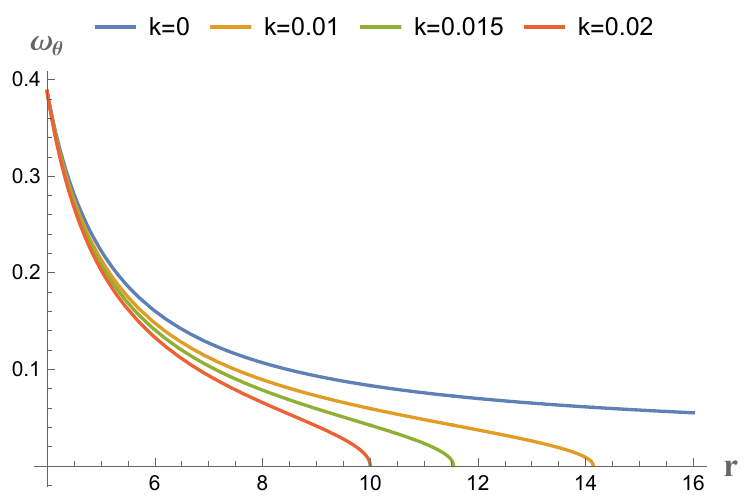}
\includegraphics[width=0.45\textwidth]{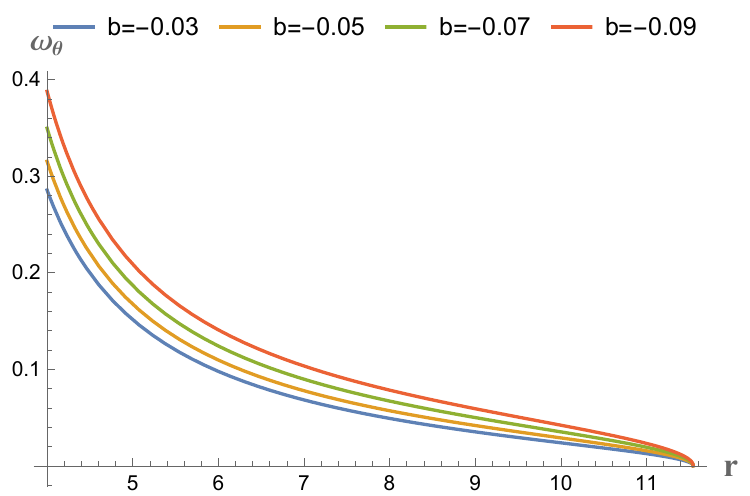}
\caption{The frequency $\omega_{\theta}$ as a function of $r$, for different values of $k$ and $b=-0.09$ ({\it Left  panel}) and for different values of $b$ and $k=0.015$ ({\it Right panel}). The other numerical values are: $M=1$  and $w=-2/3$.}
\label{omtbm}
\end{figure}

\begin{figure}[H]
\centering
\includegraphics[width=0.45\textwidth]{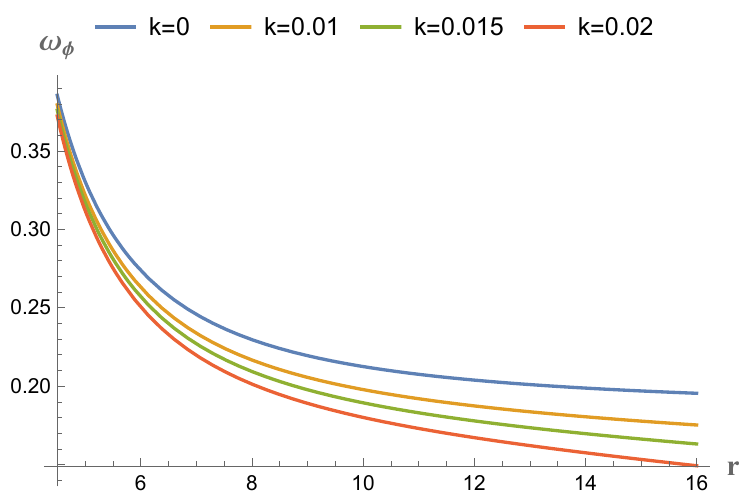}
\includegraphics[width=0.45\textwidth]{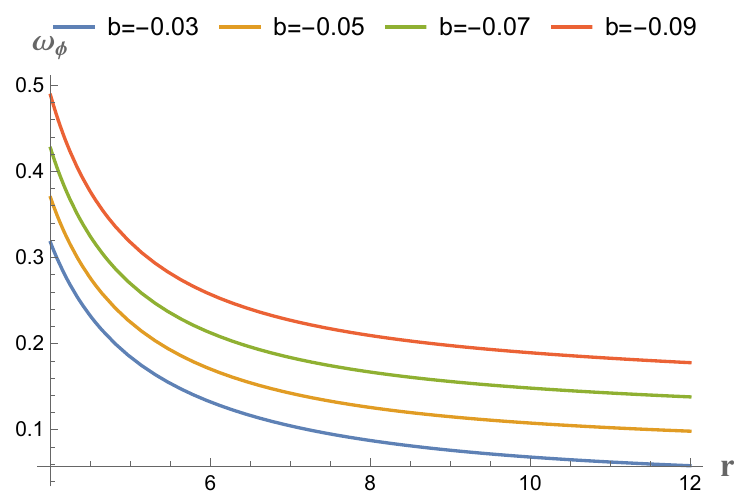}
\caption{The frequency $\omega_{\phi}$ as a function of $r$, for different values of $k$ and $b=-0.09$ ({\it Left  panel}) and for different values of $b$ and $k=0.015$ ({\it Right panel}). The other numerical values are: $M=1$  and $w=-2/3$.}
\label{omfbm}
\end{figure}

Secondly, let us mention that the same analysis can be done for $b<0$. Since the magnetic field and angular
momentum have opposite signs, the Lorentz force is attracting the charged particle towards
the $Oz-$axis. In the right panels of the figures \ref{omrbm}, \ref{omtbm} and \ref{omfbm}, one may see that all frequencies are increasing as $|b|$ increases. 

In the presence of quintessence, similarly to the previous case, all frequencies are decreasing as $k$ increases (see the left panels of the figures \ref{omrbm}, \ref{omtbm}, \ref{omfbm}). For $k=0$, both $\omega_r$ and $\omega_{\phi}$ are approaching the Larmor
frequency $\omega_L$, the first one from below and the other from above. However, by comparing the plots in the left panels of the figures \ref{omR} and \ref{omrbm}, we notice that, for $b<0$, the values of $\omega_r$ and the ranges of $r$ are much smaller since the charged particle is attracted towards the black hole by both gravitation and Lorentz force.

Also, by comparing the plots in the right panels of the figures \ref{omtf} and \ref{omfbm}, we see that, for $b<0$, the frequency $\omega_{\phi}$ is always positive and its values are significantly increased. 
 
\subsection{Prograde and retrograde motion and gravitational Larmor precession}

\subsubsection{Periapsis shift}

\begin{figure}[H]
\centering
\includegraphics[width=0.45\textwidth]{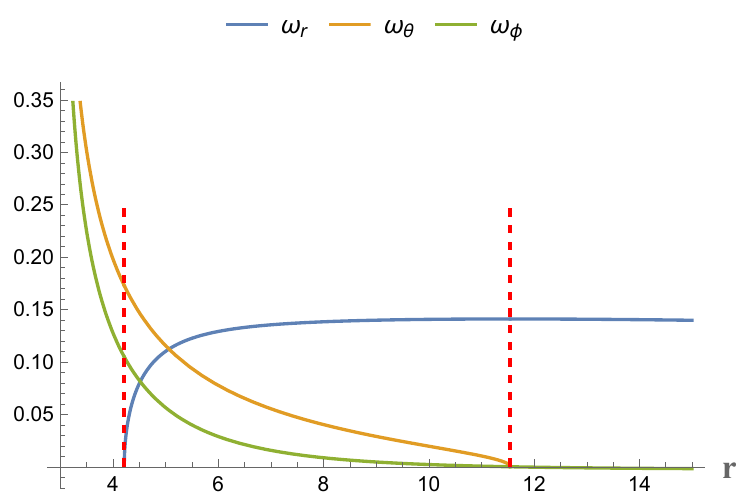}
\includegraphics[width=0.45\textwidth]{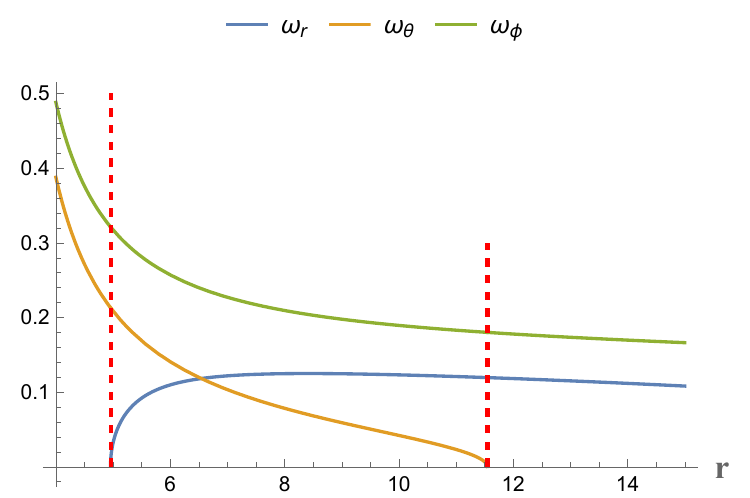}
\caption{The expressions of $\omega_r$, $\omega_{\theta}$ and $\omega_{\phi}$ as functions of $r$. The numerical values are: $M=1$, $w=-2/3$, $k=0.015$ and $b=0.09$ in the left panel and $b=-0.09$ in the right panel. The two vertical dashed red lines correspond to $r_{ISCO}$ and $r_*$.}
\label{omegas}
\end{figure}

Let us put everything together and represent the frequencies $\omega_r$, $\omega_{\theta}$ and $\omega_{\phi}$ as functions of $r$, for fixed values of $k$ and $b$.
In the left panel of the figure \ref{omegas} which corresponds to $b>0$, one may notice the special values of the radius where $\omega_r = \omega_{\phi}$ and $\omega_r = \omega_{\theta}$. 
Because the frequencies $\omega_{\theta}$ and $\omega_{\phi}$ are monotonically falling to zero for $r=r_*$, it follows that there is always an unique value of $r$ for each coincidence.
Close to ISCO where $\omega_r$ vanishes, one has $\omega_r \ll \omega_{\phi}$, while for $r$ close to $r_*$ we have $\omega_r \gg \omega_{\phi}$.
This fact has important astrophysical consequences on the periastron shift defined as
\begin{equation}
\Delta \phi = 2 \pi \left[ \frac{\omega_{\phi}-\omega_r}{\omega_r} \right]
\label{shift}
\end{equation}
whose value and sign depend on the location of the circular orbit in the allowed range $r \in [ r_{ISCO} , r_* ]$. The trajectory very close to $r_{ISCO}$ has a positive $\Delta \phi$ being a prograde one, while the trajectory close to $r_*$ is always a retrograde one. 

The same analysis can be done for negative values of $b$. For a comparison, in the the right panel of the figure \ref{omegas}, we have changed the sign of $b$ but kept the same values of the other parameters. As a result of the attractive Lorentz force, the plot of $\omega_{\phi}$ have moved upwards and the one of $\omega_r$ downwards. Since the plot of $\omega_{\phi}$ is above the one of $\omega_r$, it means that all orbits with $r \in [ r_{ISCO} , r_*)$ have a prograde precession. Once $k$ is decreasing, the two curves come closer to each other, approaching $\omega_L$, one from below and the other from above (see the left panel of figures \ref{omrbm} and \ref{omfbm}). Even though $\Delta \phi$ is decreasing for $r$ moving from $r_{ISCO}$ to $r_*$, the periastron shift remains positive for all the trajectories.

\begin{figure}[H]
\centering
\includegraphics[width=0.45\textwidth]{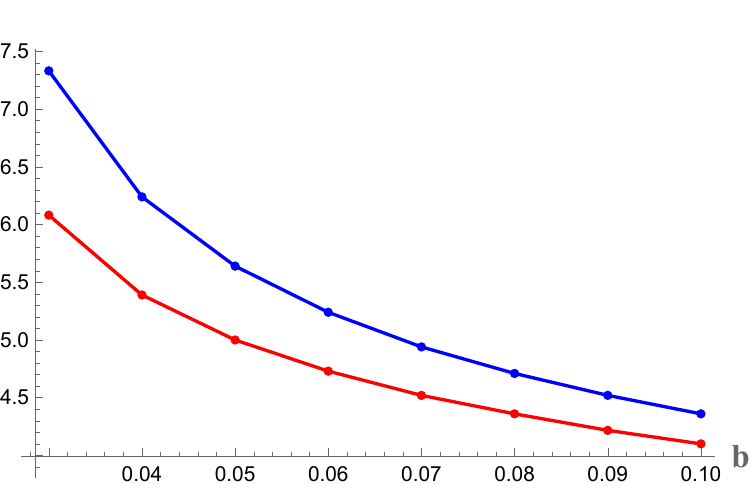}
\includegraphics[width=0.45\textwidth]{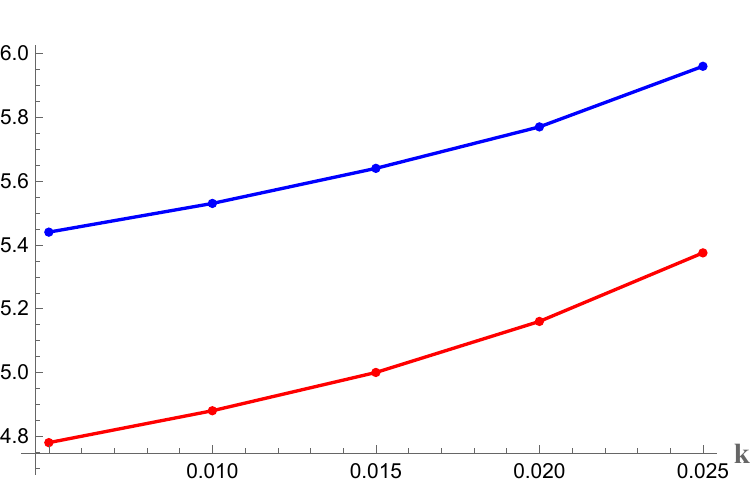}
\caption{{\it Left panel}. $r_{ISCO}$ (the red line) and $r_p$ (the blue line) as a functions of $b$ for $k=0.015$. {\it Right panel}. $r_{ISCO}$ (the red plot) and $r_p$ (the blue plot) as functions of $k$ for $b=0.05$. The other numerical values are: $M=1$ and $w=-2/3$.}
\label{Ap}
\end{figure}

For the general analysis of the periapsis shift, one may introduce the quantity
$A_p = \omega_r/\omega_{\phi}$ so that the relation (\ref{shift}) becomes
\begin{equation}
\Delta \phi = 2 \pi \left[ \frac{1}{A_p} -1 \right]
\label{shiftA}
\end{equation}
Obviously, the expression of $A_p$ is meaningful only in the region of timelike orbits stable with respect to a perturbation
in the $r$ direction, i.e. for $r \in [ r_{ISCO} , r_* ]$.
For $A_p<1$, the periapsis shift is positive, while for $A_p>1$, one obtains $\Delta \phi<0$.
Let us denote by $r=r_p$ the radius where $A_p=1$ and consequently $\Delta \phi$ is changing sign. As it can be noticed in the left panel of the figure \ref{Ap}, for $b>0$ and a given $k$, the values of $r_{ISCO}$ and $r_p$ are decreasing as $b$ is increasing. For example, for $k=0.015$, these are decreasing from $r_{ISCO} =6.08 M$ and $r_p = 7.33 M$ for $b=0.03$ to $r_{ISCO} = 4.1 M$ and $r_p =4.36M$ for $b=0.1$. The blue plot separates the region with $\Delta \phi >0$ (between the red and blue curves) from the one with $\Delta \phi <0$ (above the blue plot). One may notice that the orbit with $r=5.5M$ which was prograde for $b=0.05$ has become a retrograde one for $b \geq 0.06$. On the other hand, once $b$ is fixed, both $r_{ISCO}$ and $r_p$ are increasing with $k$ and a retrograde orbit may become a prograde one as $k$ is increasing (see the right panel of the figure \ref{Ap}). 

As we previously discussed, when the parameter $b$ is negative, the values of $r_{ISCO}$ are decreasing with $|b|$ and $\Delta \phi$ keeps the positive sign.

\subsubsection{The gravitational Larmor precession}

If the
orbit is slightly inclined to the equatorial plane, a gravitational Larmor precession may appear \cite{Chakraborty:2022yso}. This is characterized by the difference between $
\omega_{\phi}$ and $\omega_{\theta}$ as
\begin{equation}
\Delta \theta = 2 \pi \left[ \frac{\omega_{\phi} - \omega_{\theta}}{\omega_{\theta}} \right] = 2 \pi \left[ \frac{1}{A_n} -1 \right]
\label{nod}
\end{equation}
where $A_n = \omega_{\theta} / \omega_{\phi}$.
This motion is equivalent to the so-called LT precession that was first derived by Lense and Thirring for the
slowly rotating Kerr spacetime and later by Kato.
Even though it was assumed that the expressions
$\omega_{\phi}$ and $\omega_{\theta}$ match each other in any
non-rotating spacetime, recently, it has been shown that
the gravitational Larmor precession,
which is similar to the LT effect in the Kerr
spacetime, can arise when in the Schwarzschild black hole is immersed in a magnetic field \cite{Chakraborty:2022yso}.
In view of the formulas (\ref{omegaf}) and (\ref{omt}), the expression of $A_n$ has the explicit form
\begin{equation}
A_n = \sqrt{ \frac{L+br^2}{L-br^2}}
\label{An}
\end{equation}
where $L> |b| r^2$.
For $b>0$, one has $A_n >1$ and $\Delta \theta <0$, while for $b<0$, one has $A_n <1$ and $\Delta \theta >0$.

This can be also seen in the figures \ref{omegas} where, for $b>0$, the plot corresponding to $\omega_{\phi}$ is below the one of $\omega_{\theta}$ and therefore $\Delta \theta < 0$ for $r \in [ r_{ISCO} , r_*]$. The two frequencies are monotonically decreasing and they are vanishing in $r=r_*$. On the other hand since, for $b<0$, $\omega_{\phi}$ is above the plot of $\omega_{\theta}$, it follows that $\Delta \theta$ keep a positive sign.

Conclusions of the discussion on the signs of $\Delta \phi$ and $\Delta \theta$ presented in this section are given in table \ref{table3}, for positive and negative values of $b$ and $r$ in its allowed range $r \in [r_{ISCO} , r_p]$. 
 
 \begin{table}[h!]
\centering
\begin{tabular}{|c|c|c|}
\hline
 $r$ & $b>0$ & $b<0$ \\ \hline
$r \in ( r_{ISCO} , r_p]$ & $\Delta \phi  \geq 0$ & $\Delta \phi \geq 0$  \\ \hline
$r \in ( r_p , r_*]$ & $\Delta \phi <0$ & $\Delta \phi >0$  \\ \hline
$r \in [r_{ISCO} , r_* )$ & $\Delta \theta <0$ & $\Delta \theta >0$  \\ \hline
\end{tabular}
\caption{The signs of $\Delta \phi$ and $\Delta \theta$ for $r \in [ r_{ISCO} , r_* ]$.}
\label{table3}
\end{table}

\subsection{Illustrative examples of quasi-circular orbits}

\begin{figure}[H]
    \centering
    \begin{subfigure}{0.4\textwidth}
        \centering
        \includegraphics[scale=0.5, trim=0cm 7cm 0cm 2cm, clip]{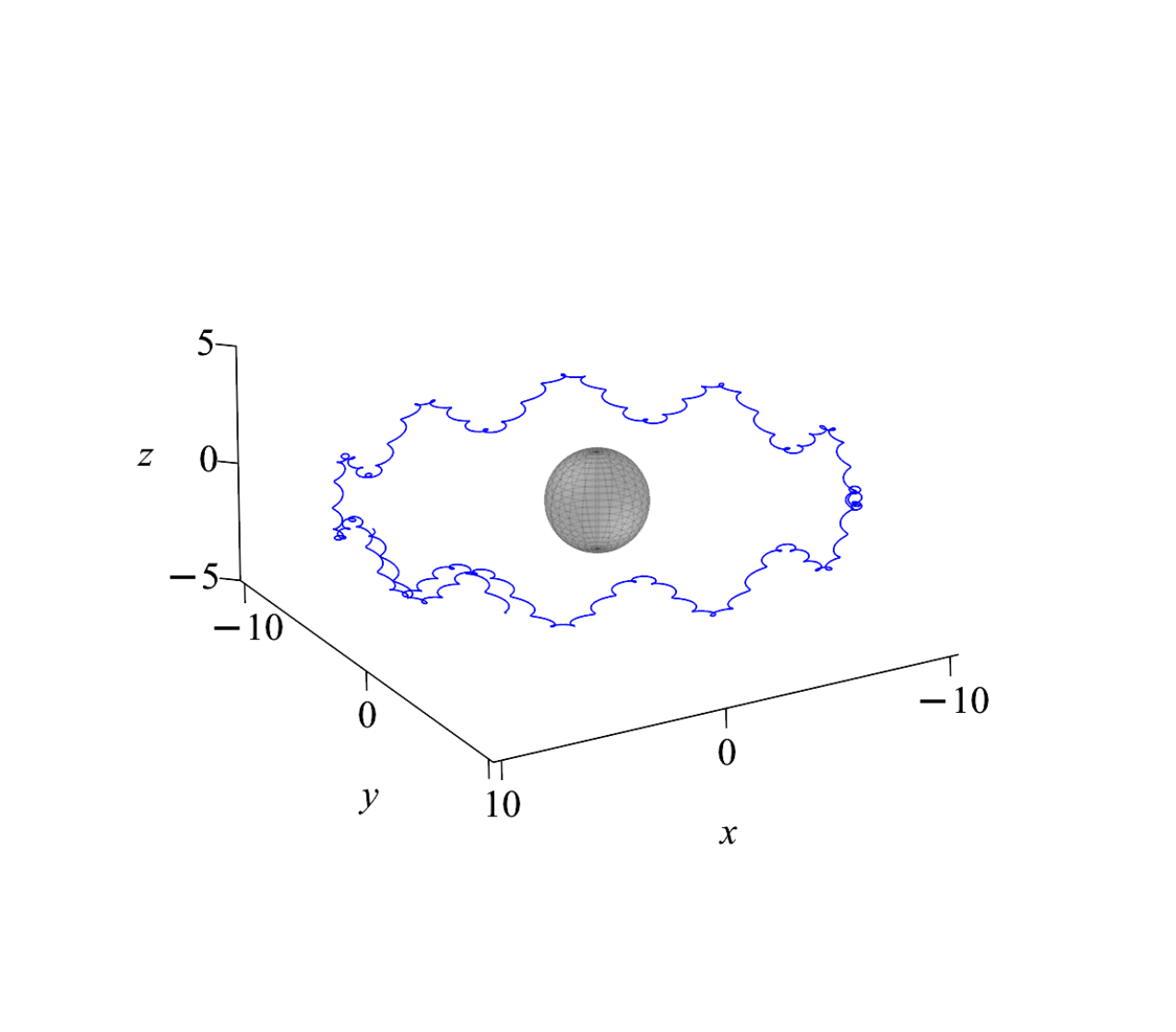}
    \end{subfigure}
    \hfill
    \begin{subfigure}{0.52\textwidth}
        \centering
        \includegraphics[scale=0.55, trim=0cm 12cm 0cm 2cm, clip]{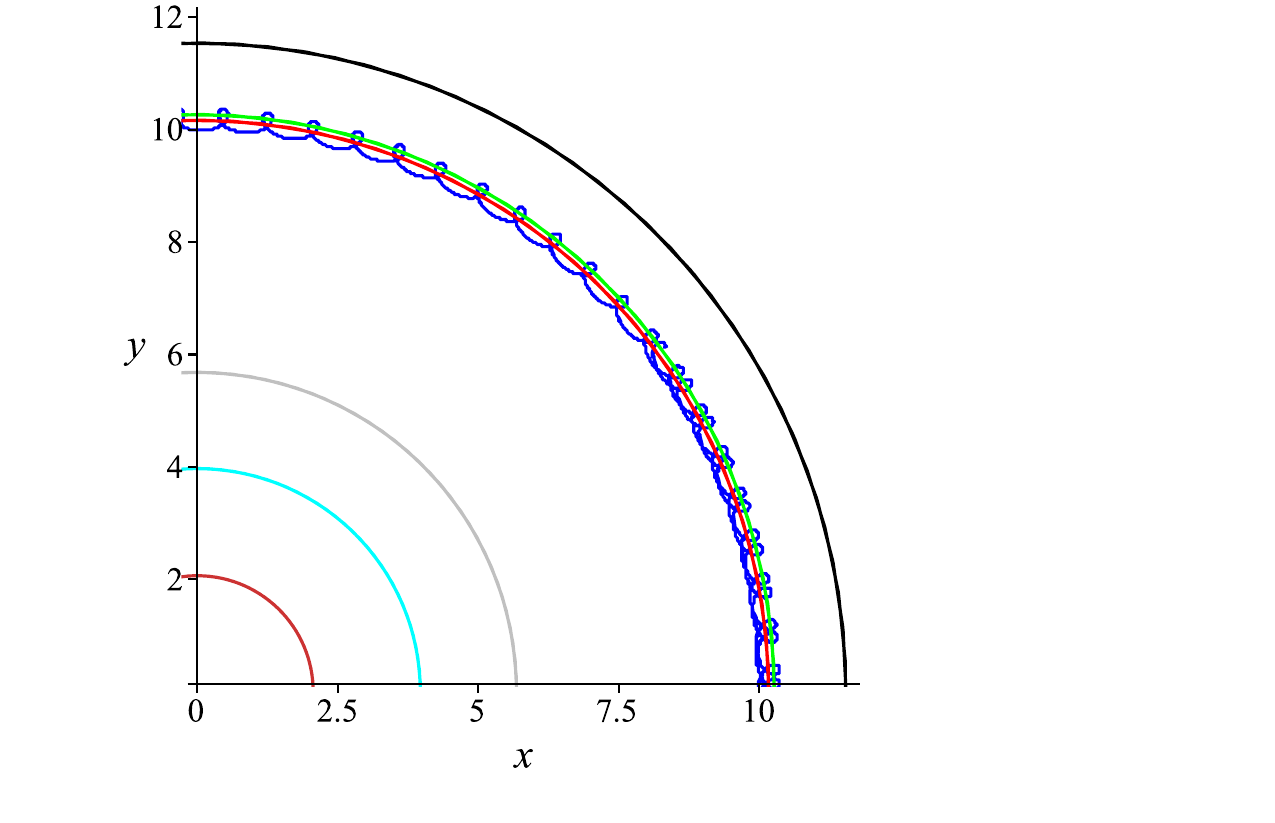}
    \end{subfigure}
\caption{A 3D representation of a stable quasi-circular trajectory with $r_c$ close to $r_*$ and a quarter of its projection on the $xOy$ plane. The numerical values are: $M=1$, $w=-2/3$, $k=0.015$, $b=0.09$, $L=9.5$, $E^2=0.652$. The dark red circle is the horizon $r_- =2.064$, the cyan circle is the ${ISCO}$ with $r_{ISCO}=4.22$, the gray circle has $r_p =5.68$ which corresponds to the change of sign of $\Delta \phi$, the red circle is the unperturbed circular orbit with $r_c = 10.17$, the green circle of radius $r_0 =10.274$ is separating the region of curly trajectories from the non-curly one and the black circle corresponds to $r_* =11.547$.}
\label{circ}
\end{figure}

As a first example of a perturbed circular orbit of radius $r_c$, let us consider the stable trajectory with $r_c$ close to $r_* = \sqrt{2M/k}$ represented in the figure \ref{circ}. In the three-dimensional representation given in the left panel, one may notice the quasi-harmonic oscillations both around $r_c$ and with respect to the equatorial plane. In the projection on the equatorial plane (a quarter is given in the right panel), the region of stable circular orbits is between the cyan and black circles which correspond to $r_{ISCO}$ and $r_*$ respectively. The values of $L$ and $E^2$ are in the shaded regions represented in the figures \ref{fig:Lreg} and \ref{fig:Ereg}. The unperturbed circular orbit of radius $r_c$ is represented by the red circle which is very close to the green circle of radius $r_0 = \sqrt{L/b}$ where $\phi$ is changing the sign. As expected, because $r_c >r_p$, the orbit has a negative periastron shift and is curly since one of the turning points is beyond $r_0$. 

\begin{figure}[H]
    \centering
    \begin{subfigure}{0.4\textwidth}
        \centering
        \includegraphics[scale=0.55, trim=0cm 7cm 0cm 2cm, clip]{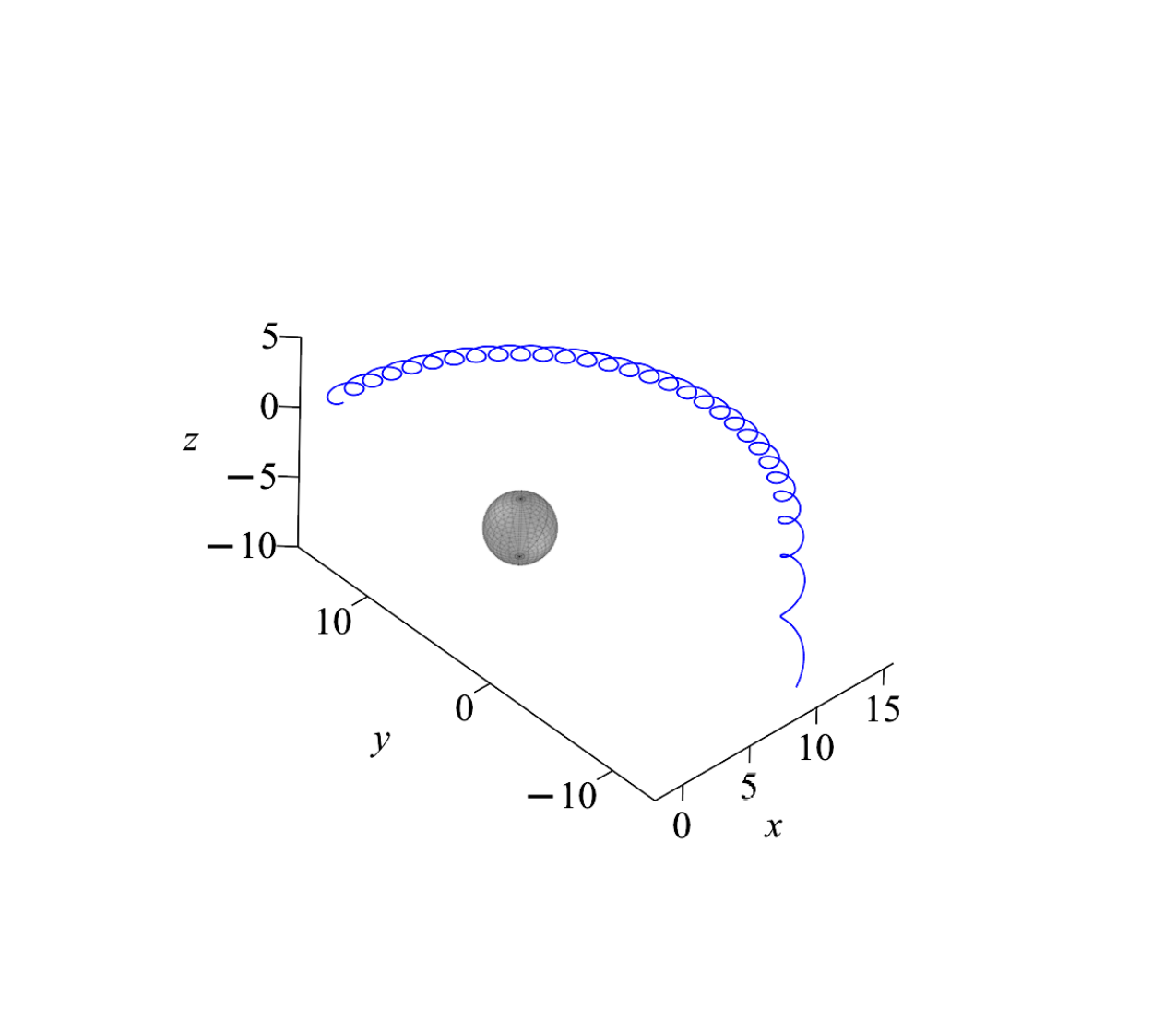}
    \end{subfigure}
    \hfill
    \begin{subfigure}{0.52\textwidth}
        \centering
        \includegraphics[scale=0.55, trim=0cm 12cm 0cm 2cm, clip]{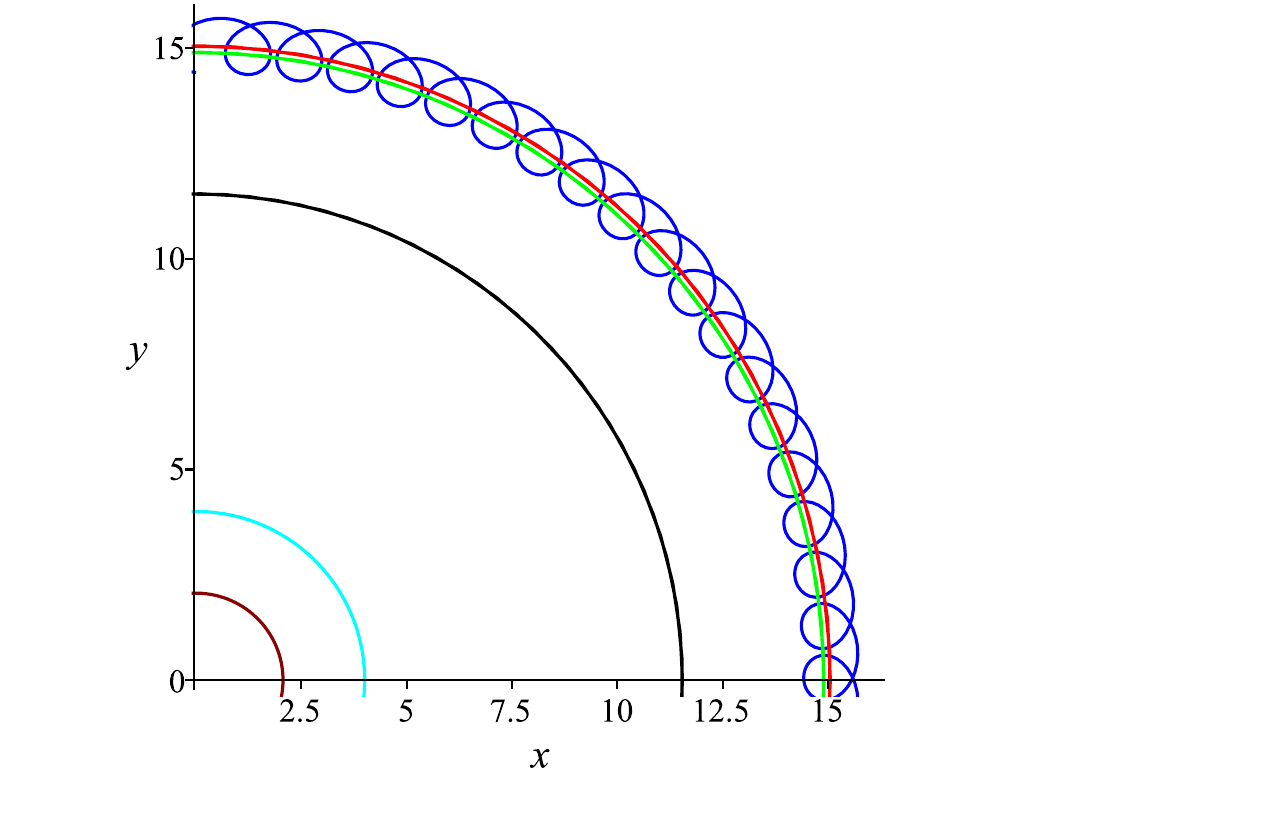}
    \end{subfigure}
\caption{A 3D representation of an vertically unstable quasi-circular trajectory with $r_c>r_*$ and its projection on the $xOy$ plane. The numerical values are: $M=1$, $w=-2/3$, $k=0.015$, $b=0.09$, $L=20$, $E^2=0.65$. The dark red circle is the horizon $r_- =2.064$, the cyan circle is the ${ISCO}$ with $r_{ISCO}=4.22$, the black circle corresponds to $r_* =11.547$, the green circle of radius $r_0 =14.90$ is separating the region of curly trajectories from the non-curly one and the red circle is the unperturbed circular orbit with $r_c = 15.06$.}
\label{fig:circu}
\end{figure}

In view of the analysis developed in section 5, let us mention that circular orbits with $r_c >r_*$ are unstable. Such an example is given in the figure \ref{fig:circu} where one may also notice that the orbit of the particle escaping towards the cosmological horizon along the $z$-axis is curled toward the black hole. This happens because the repulsive quintessence contribution becomes dominant for $r>r_*$. On the other hand, in figure \ref{circ}, we see an opposite behaviour for the particle's stable orbit with radius $r<r_*$ which is curled outward the black hole as a result of gravity dominance.

\begin{figure}[H]
    \centering
    \begin{subfigure}{0.4\textwidth}
        \centering
        \includegraphics[scale=0.5, trim=0cm 7cm 0cm 2cm, clip]{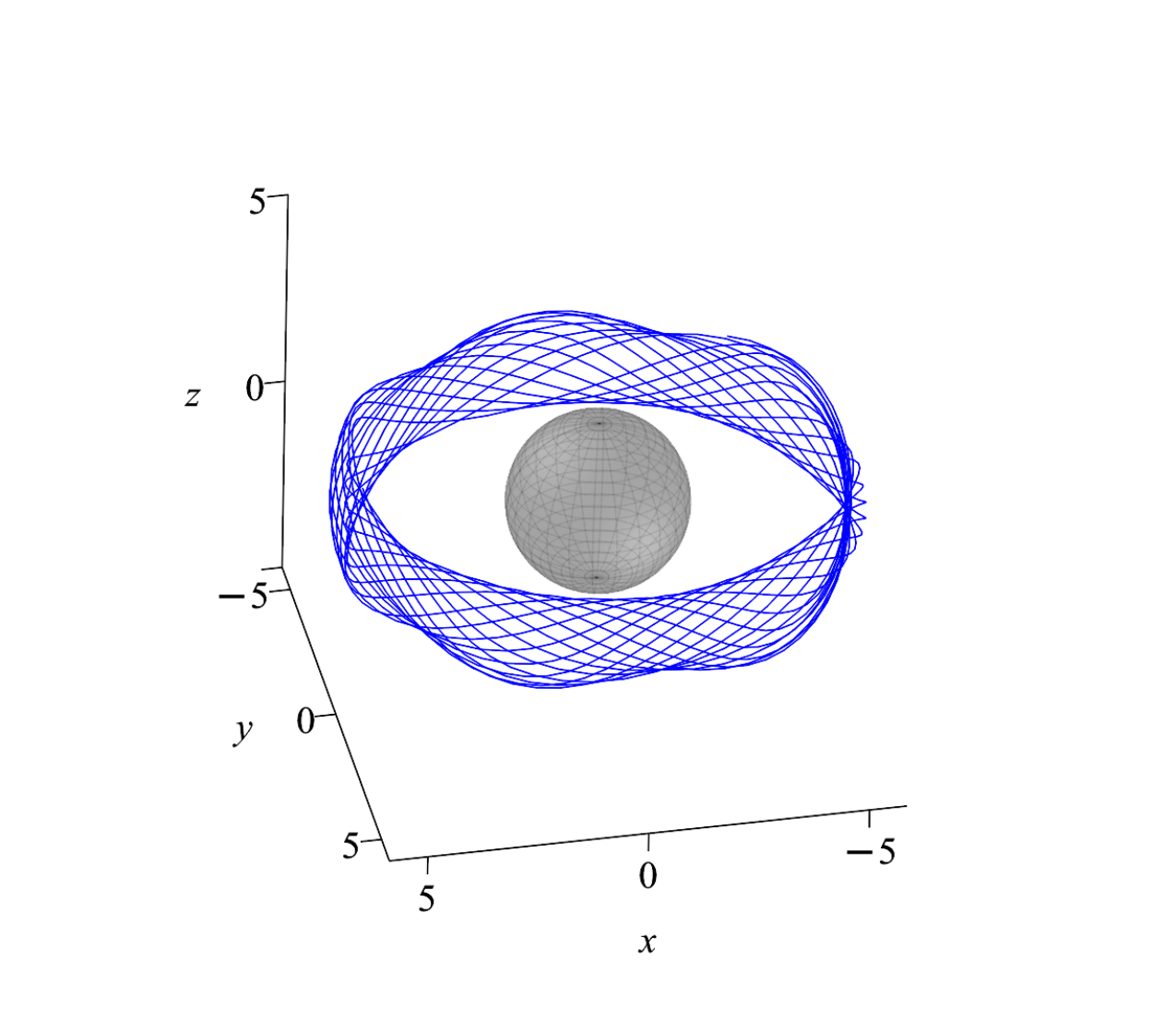}
    \end{subfigure}
    \hfill
    \begin{subfigure}{0.52\textwidth}
        \centering
        \includegraphics[scale=0.55, trim=0cm 12cm 0cm 2cm, clip]{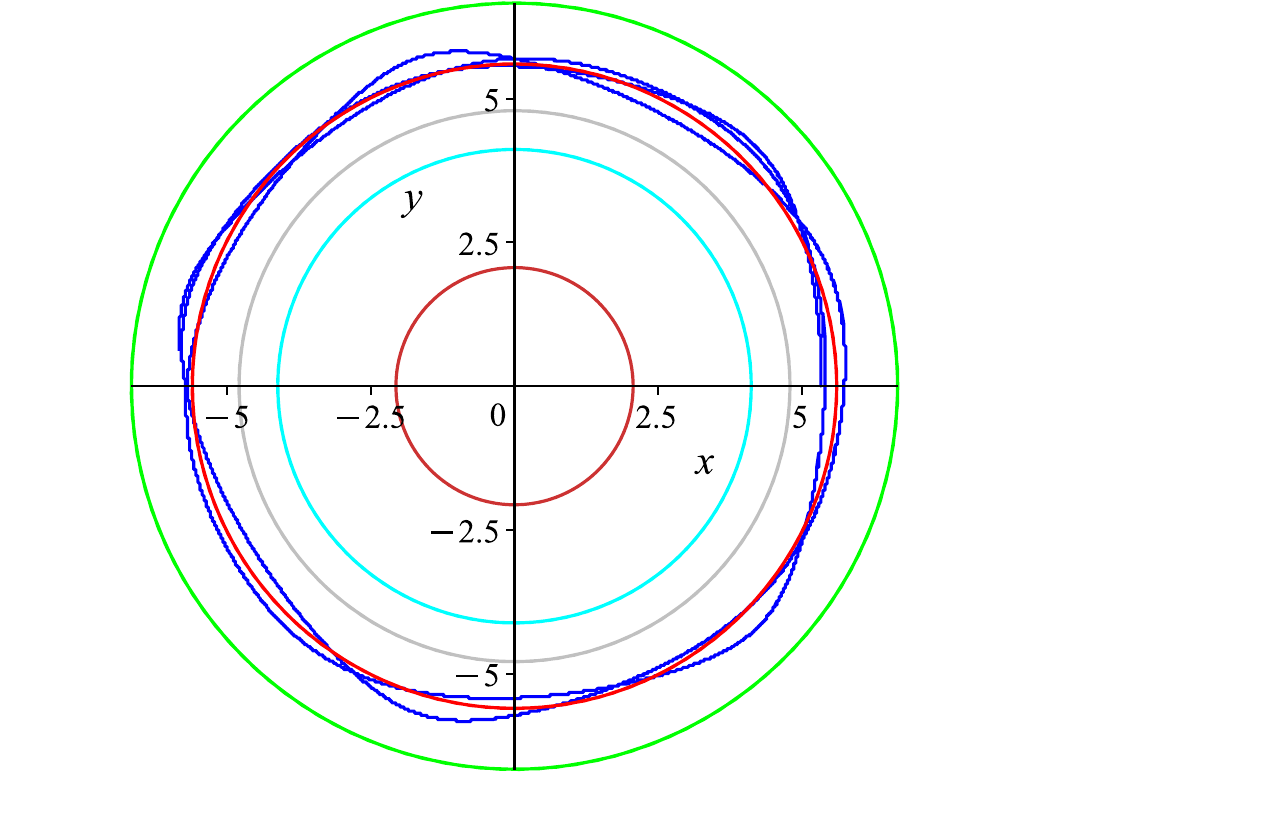}
    \end{subfigure}
\caption{A 3D representation of a stable quasi-circular trajectory with $r_c$ between $r_*$ and $r_{ISCO}$ and its projection on the $xOy$ plane. The numerical values are: $M=1$, $w=-2/3$, $k=0.015$, $b=0.09$, $L=4$, $E^2=0.59$. The dark red circle is the horizon $r_- =2.064$, the cyan circle is the ${ISCO}$ with $r_{ISCO}=4.22$, the gray circle corresponds to the change of sign of $\Delta \phi$, the red circle is the unperturbed circular orbit with $r_c = 5.60$, the green circle of radius $r_0 =6.66$ is separating the region of curly trajectories from the non-curly one.}
\label{circ0}
\end{figure}

In the case where the radius of the unperturbed stable circular orbit is approaching $r_{ISCO}$, non-curly trajectories as the one represented in the figure \ref{circ0} exist for suitable values of $L$ and $E$. Since $r_p<r_c$, the periastron shift is negative.

\begin{figure}[H]
    \centering
    \begin{subfigure}{0.49\textwidth}
        \centering
        \includegraphics[scale=0.5, trim=0cm 7cm 0cm 2cm, clip]{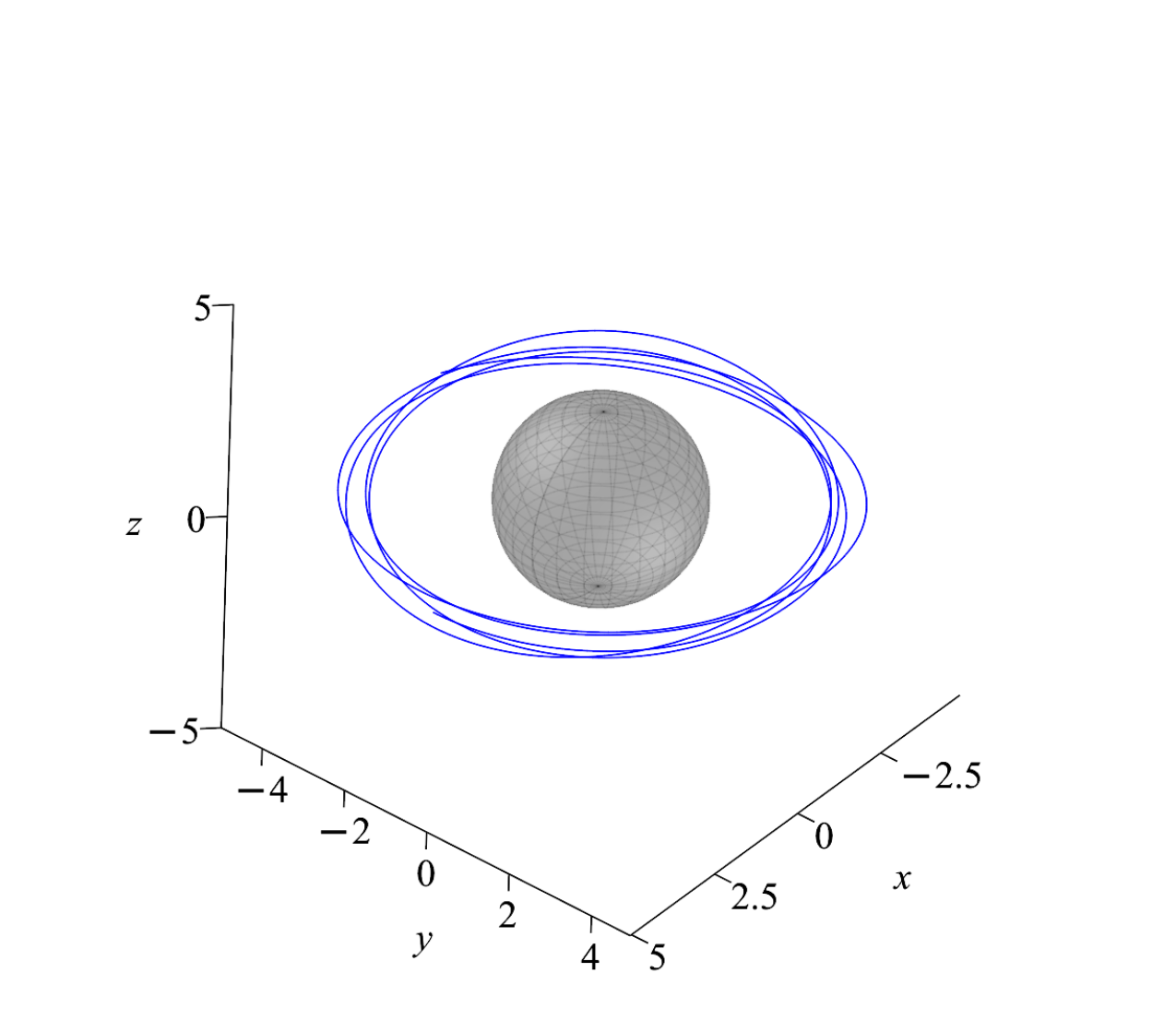}
    \end{subfigure}
    \hfill
    \begin{subfigure}{0.49\textwidth}
        \centering
        \includegraphics[scale=0.5, trim=0cm 12cm 0cm 2cm, clip]{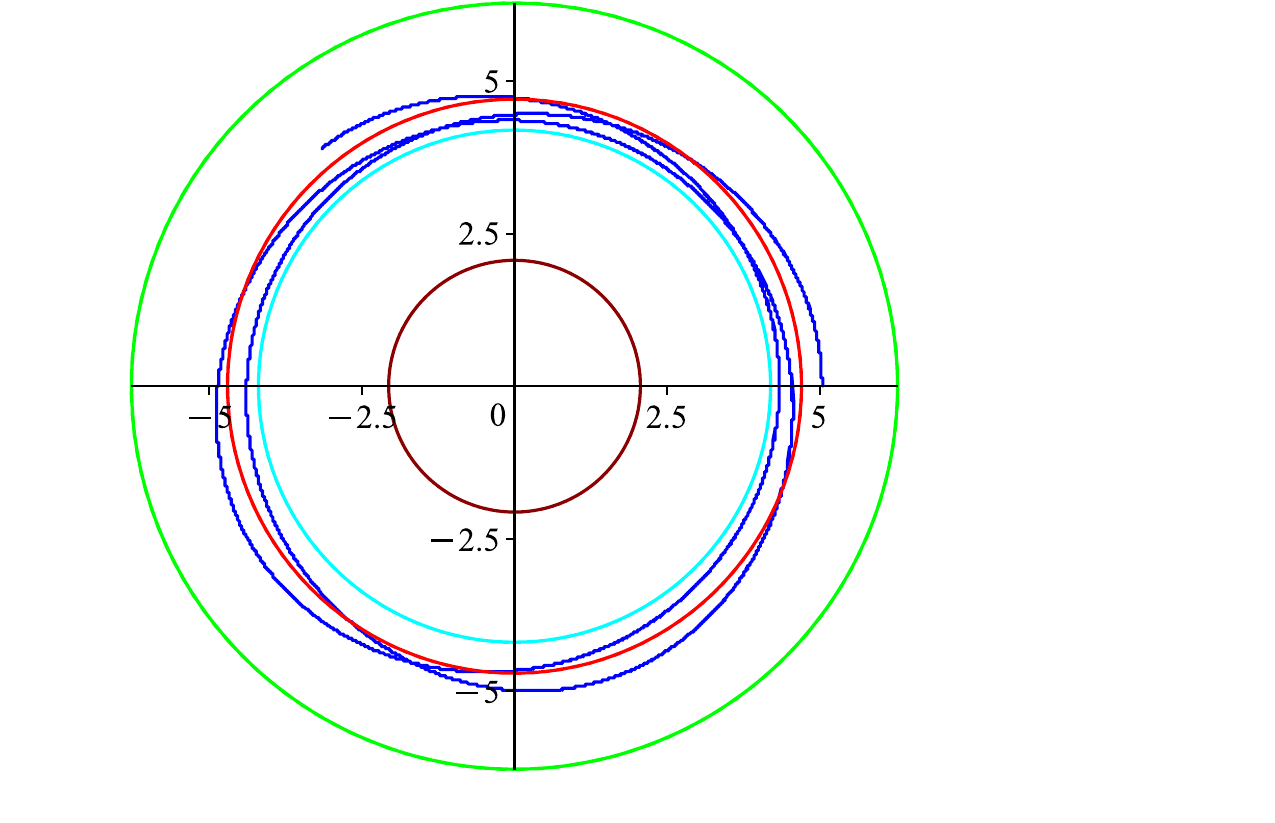}
    \end{subfigure}
\caption{A 3D representation of a stable quasi-circular trajectory with $r_c$ close to $r_{ISCO}$ and its projection on the $xOy$ plane. The numerical values are: $M=1$, $w=-2/3$, $k=0.015$, $b=0.09$, $L=3.55$, $E^2=0.561$. The dark red circle corresponds to the horizon $r_-= 2.064$, the cyan circle represents the $ISCO$ with $r_{ISCO} = 4.22$, the red circle is the unperturbed trajectory with $r_c= 4.70$, the green circle has $r_0 = 6.28$. }
\label{circ1}
\end{figure}

Finally, as a physically important example, in the figure \ref{circ1}, we have considered the perturbed stable circular orbit with $r_c$ close to $r_{ISCO}$. Similarly to the previous case, the trajectory is non-curly since it does not cross the blue circle of radius $r_0 = \sqrt{L/b} = 6.28$ separating the region of curly trajectories from the non-curly ones.

\section{Conclusions}

From an astrophysical perspective, our investigation may serve as an effective framework for describing the motion of charged particles in realistic environments where supermassive or stellar-mass black holes surrounded by quintessence are also immersed in large-scale magnetic fields. Observations from type Ia supernovae \cite{SupernovaCosmologyProject:1998vns} and microwave background radiation \cite{WMAP:2006bqn} indicate that the Universe expands at an accelerated rate, being driven by the mysterious dark energy characterized by a negative pressure.
Numerous theoretical models have been proposed to explain the nature of the dark energy and one of the most prominent candidate is the quintessence.
For realistic cosmological scenarios, the energy density of the quintessential fluid is expected to be very small \cite{Cao:2023ppv, Dubey:2025cdk}. However, it alters the structure of the effective potential at large distances, modifies the stability conditions and influences particle trapping, escape channels and large-scale disk morphology. 
Observationally, the quantities most sensitive to the combined magnetic and quintessence effects are the ISCO radius and the associated periastron and nodal precessions. Moreover, the the radial and vertical epicyclic frequencies are directly connected to high-frequency quasi-periodic oscillations (HFQPOs) observed in X-ray binaries and to disk precession phenomena in accreting systems \cite{Aliev:2002nw}. 

The paper is based on the metric (\ref{metmag}) which is describing a magnetized black hole surrounded by quintessential fluid, as a generalization of the Kiselev solution \cite{Kiselev:2002dx} in the presence of an uniform magnetic field. We do not provide details of the procedure of obtaining this solution, since these are given in our paper \cite{Lungu:2024iob}. In the absence of quintessence, i.e. $k=0$, the metric turns into the well-known Ernst solution \cite{Ernst:1976mzr} while for $B_0 =0$, we obtain the Kiselev spacetime.
The weak magnetic field approximation which is considered in our work is suitable for real astrophysical situations. Indeed, the estimated values of the magnetic induction for supermassive black holes \cite{Daly:2019srb} show that one can neglect the back reaction of the external magnetic field on the space-time geometry. However, the Lorentz force acting on charged particles has a significant influence on their trajectories whose type and shape depend on the model's parameters $k$, $w$ and $B_0$.

Even though configurations of magnetic fields in the background of Schwarzschild spacetime have been intensively studied \cite{Frolov:2010mi, Lim:2015oha}, the presence of quintessential matter surrounding the black hole leads to new interesting features. For example, besides the black hole's horizon, it exist a cosmological horizon so that the allowed range of the radial coordinate is $r \in [r_- , r_+]$. For suitable values of the model's parameters $b = \varepsilon B_0$ and $k$, we have derived regions of the radial coordinate that host stable circular orbits. This result may have measurable consequences for the detection of quintessential matter around magnetized black holes or/and for determining the magnetic field strength.

Once the effective potential is known, one can establish what types of orbits are possible. Contrary to Ernst spacetime, the potential (\ref{V}) is vanishing on the horizons and it possesses off-equatorial critical points which correspond to saddle points, this feature being a result of quintessence.
For a bound orbit to exist, the condition (\ref{BoundCond}) should be satisfied and this leads to a range of the angular momentum given in (\ref{Lrange}) and to a maximum value of the radial coordinate of the turning point, defined in (\ref{rmax}), which is depending only on the quintessence parameters $k$ and $w$. 
Also, in the figures \ref{fig:Lreg} and \ref{fig:Ereg}, the key features are the shaded regions which correspond to the values of $L$ and $E$ which allow bound orbits. 

A special attention is given to circular orbits. Since the potential vanishes on the horizons and it possesses at least one maximum value inbetween, it always exists un unstable circular orbit. On the other hand, stable circular orbits are possible for small values of $k$, lower than a critical value, and for the magnetic parameter in the range $b_{min} < b < b_{max}$, with $b_{min}$ and $b_{max}$ given in (\ref{bmin}) and (\ref{bmax}). Outside this range whose width is decreasing as $k$ increases, no stable circular orbits exist. The condition $b_{min} = b_{max}$ is satisfied for $r=r_*$, where $r_*$ is given in (\ref{rmax}). As we pointed out, the static radius $r=r_*$ is particularly important since it separates the region of stable orbits from the one where only unstable trajectories exist. The constrains on the quintessence parameter $k$ and the $r_{ISCO}$ ranges are collected in table 3. An important feature is that, due to quintessence, values of $r_{ISCO}$ bigger than the reference value $r=6M$ are possible. However, once the magnetic field comes into place, the ISCO radius shifts to lower values approaching the limiting value $r=3M$ (see the figure \ref{fig:breg}).

As a main goal, in section 6, we address the problem of perturbed quasi-circular orbits. 
In this respect, we use the conditions of existence of stable circular orbits discussed in the previous section and focus on the 
fluctuations about these orbits and their stability expressed by the reality of the radial and transversal frequencies. 
For suitable values of $L$ and $E$ and $r \in [ r_{ISCO} , r_* ]$, one may find circular trajectories which are stable both under fluctuations in the equatorial plane and against fluctuations in the transverse direction, while for $r>r_*$, the quintessence destabilizes any bound orbit.
As expected, once the quintessence parameter is increasing, the characteristic frequencies are decreasing and there is a critical value of $k$ above which the trajectory becomes unstable. 
On the other hand, the values of the frequencies and the allowed range of $r$ are depending on the sign and values of the magnetic parameter. For example, for $L>0$ and $b < 0$, the Lorentz force is attracting the charged particle towards the Oz-axis and all frequencies are increasing with $|b|$.

In addition, we discuss the periastron and nodal precessions. The values and signs of $\Delta \phi$ and $\Delta \theta$ are depending not only on the magnetic parameter $b$, but also on the quintessence parameter $k$. As an interesting result, as $k$ increases, retrograde orbits may turn into prograde ones.

In the final subsection 6.3, we are giving some illustrative examples of trajectories for specific choices of the model's parameters.
In order to draw the particle's trajectory by solving the equations of motion (\ref{weak}) using a numerical procedure, one must verify all the conditions previously discussed and fix suitable values for $B_0$, $k$ and $L$. The coloured circles in the projections of the trajectories on the $xOy$ plane (see the figures \ref{circ}--\ref{circ1})  are dividing physically important regions, as for example the green circle separates the region between curly and non-curly orbits, the gray circle corresponds to the change of sign of $\Delta \phi$, while the black circle corresponds to $r_*$ above which no stable trajectory exists. Similarly to Ernst spacetime, the particle's trajectory may be curly or non-curly \cite{Lim:2015oha}. However, when the quintessence contribution is dominant, the unstable trajectory is curled toward the black hole and this behavior has no equivalence in Ernst spacetime where the trajectory is always curled outward the black hole.

\section*{Acknowledgements}
The authors would like to thank the anonymous Referee whose remarks and suggestions helped us to improve our manuscript.

\end{document}